\documentclass[11pt,a4paper]{article}
\usepackage{jheppub}
\usepackage{subcaption}
\usepackage{caption}

\usepackage{mathtools}
\DeclareMathOperator{\ReN}{Re}

\DeclareMathOperator{\tr}{Tr}
\newcommand{\dif}[1]{\mathrm{d} #1 }
\newcommand{\effAction}{S^{\text{eff}}}
\newcommand{\pureGaugeEffAction}{\effAction_{\text{G}}}
\newcommand{\quarkEffAction}{\effAction_{\text{Q}}}

\usepackage{tikz}
\usepackage{pgfplots} 
\usepackage{pgfgantt}
\usepackage{pdflscape}
\pgfplotsset{compat=newest} 
\pgfplotsset{plot coordinates/math parser=false}
\newlength\fwidth

\title{From deconfinement to nuclear matter: mean-field approaches for effective Polyakov loop theories of lattice QCD}
\author[a]{Christoph Konrad,}
\author[a]{Owe Philipsen}
\affiliation[a]{Institute for Theoretical Physics, Goethe-University Frankfurt am Main, Max-von-Laue-Str. 1, 60438 Frankfurt am Main, Germany}
\emailAdd{konrad@itp.uni-frankfurt.de}
\emailAdd{philipsen@itp.uni-frankfurt.de}

\abstract{Numerical studies of the QCD phase diagram at finite baryon chemical potential $\mu_B$ on the lattice are impeded by a sign problem. Effective Polyakov loop theories derived from lattice QCD via combined strong-coupling and hopping expansions are valid for heavy quarks only, but tractable with a significantly less severe sign problem. In this work, we apply three mean-field approximations to these effective theories, each incorporating local fluctuations to different degrees. We compare their predictions against Monte Carlo results for the deconfinement transition at high temperatures and for the baryon onset transition at low temperatures. In agreement with earlier effective theory simulations, we find a first-order nuclear liquid-gas transition with a critical end point for very low temperatures and moderately heavy quarks. While the location
of fluctuation-dominated critical end points is expectedly inaccurate, the phase diagram of lattice QCD with heavy quarks can be determined qualitatively with entirely analytical methods.    }

\begin{document}
	\maketitle
	
	\section{Introduction}
	Determining the QCD phase diagram remains a challenging research task due to the sign problem 
	in Monte Carlo  evaluations of  lattice QCD (LQCD) at non-zero baryon chemical potential $\mu_B$. To overcome this issue, effective theories of LQCD have been derived from Wilson's standard lattice action with arbitrary chemical potentials by using a combination of strong-coupling and hopping parameter expansions~\cite{Langelage2011, Langelage:2014vpa, Glesaaen2016}. These effective theories are three-dimensional with Polyakov loops as field variables. Their numerical Monte Carlo evaluation
	is computationally cheaper compared to full LQCD at $\mu_B = 0$~\cite{Bergner:2013qaa} and, because
	of a significantly reduced sign problem, 
	is also possible at $\mu_B\not = 0$ \cite{Fromm2012, Fromm2013, Langelage:2014vpa, Glesaaen2016}. Moreover,
	these effective theories are also amenable to analytic series expansion approaches in the effective couplings, thus avoiding the sign problem altogether~\cite{Langelage:2014vpa, Glesaaen2016, Philipsen:2019qqm}. Using these effective theories it was possible to obtain the critical couplings of SU(3) pure gauge theory to better than 10\% accuracy \cite{Langelage2011} and to determine the second-order critical line where the first-order deconfinement transition for QCD with heavy quarks turns into a crossover in the $\mu_B - m_\pi - T$-phase diagram \cite{Fromm2012}. In the low-temperature  regime it was possible to study the baryon onset transition to cold nuclear matter and its quark mass dependence~\cite{Fromm2013,Langelage:2014vpa}, as well as identifying a quarkyonic regime with the pressure scaling linearly with the number
of colours beyond this transition~\cite{Philipsen:2019qqm}. However, the evaluation of these effective theories is made difficult by the appearance of long-range and multi-point interactions in higher orders in the expansion parameters, which are necessary if one is interested in lighter quark masses and finer lattices.	
	
 A well-known approach for handling such complexities is the mean-field approximation, which is expected to become more accurate as the coordination number of the interactions 
 increases. It has already been applied successfully to Polyakov loop models corresponding to low-order versions of the effective actions discussed in the following~\cite{Dumitru:2005ng, Fukushima:2006uv, Greensite:2012xv,Rindlisbacher:2015pea, Borisenko:2020cjx}. In the present work our motivation is twofold: first we refine
 these previous mean-field studies by including resummations of local fluctuations, in order to render their predictions more
 accurate when compared to fully non-perturbative evaluations. We then test these techniques 
 in effective theories with a growing number of couplings, representing full LQCD with increasing accuracy and range of validity. 
	
Specifically, we apply mean-field approximations in three variations, each capturing a different amount of local fluctuations,
and test their predictions for the finite temperature deconfinement transition against Monte Carlo results, both for vanishing and non-vanishing baryon density. 
While the most
naive mean-field approach is found to only yield qualitative results for the location of the first-order transition line (about 49\% error), our best resummed approach gives quantitative results (about 1.5\% error). A second-order critical end point
is correctly predicted by all approaches, but its location is inaccurate in all cases, since mean field is well known to break down once the dynamics of a system is fluctuation driven. 
In the low-temperature regime we demonstrate the applicability of a mean-field analysis for the onset transition to baryon matter. 
Consistent with earlier investigations by complex Langevin simulations~\cite{Langelage:2014vpa}, this
transition is a crossover for any non-zero temperature and very heavy quarks, whereas it turns
into the expected first-order liquid-gas transition with a critical end point for light quarks.  
In this parameter regime the entropy behaves differently from that of the physical
liquid-gas transition, and becomes unphysical for zero temperature, in agreement with a perturbative analysis following earlier work~\cite{Langelage:2014vpa}.  This systematic error
can be attributed to the truncation of the effective theory and affects the finite temperature regime as well, albeit more mildly. 
	
	This paper is organized as follows. In section \ref{sec:Overview_effective_theories} we review the derivation of the effective theories. In section \ref{sec:General_mean_field_discussion} we discuss three variations of mean-field approximations and apply them in section \ref{sec:Evaluation}. Specifically, we study the deconfinement transition at high temperatures and the nuclear liquid-gas transition at low temperatures, where also the newly observed systematic error of the effective theories is discussed.

\section{Overview over the effective theories \label{sec:Overview_effective_theories}}

	The effective theory is derived from the partition function of LQCD after integrating out $N_f$ mass-degenerate quarks at identical chemical potentials \cite{Langelage2011, Fromm2012},
	\begin{align}
		Z &= \int[\dif{U_\mu}]e^{-S_G[U_\mu]}\left(\det Q[U_\mu]\right)^{N_f},
	\end{align}
	where $S_G$ is Wilson's gauge action and $Q$ is the Wilson-Dirac quark operator. The quark determinant is split into  static and kinetic contributions \cite{Langelage:2014vpa},
	\begin{align}
		\det Q = \det Q_{\text{S}} \det Q_{\text{K}}.
	\end{align}
	The static quark determinant $\det Q_{\text{S}}$ describes quark world lines extending in the temporal direction only and factorizes over spatial lattice sites,
	\begin{align}
		\det Q_{\text{S}} &= \prod_{\mathbf{x}}\det Q_{\text{S}}^{\text{loc}}(\mathbf{x}) = \prod_{\mathbf{x}}\det\left(\vphantom{\frac{1}{1}}1 + h_1 W_{\mathbf{x}}\right)^{2} \det\left(\vphantom{\frac{1}{1}}1 + \bar{h}_1 W_{\mathbf{x}}^\dagger\right)^{2} \label{eq:ExpressionLocalPartStaticDeterminant}\\
		h_1 &= \exp(N_\tau(a\mu + \ln(2\kappa)) + \dots),\quad \bar{h}_1(\mu) = h_1(-\mu),
	\end{align}
	where $\mu$ is the quark chemical potential and we introduced the static (anti-)quark effective coupling $h_1$ ($\bar{h}_1$). The kinetic quark determinant $\det Q_{\text{K}}$ includes spatial hops of quarks and its contributions to the effective theories will be discussed further in section \ref{sec:Introduction_kinetic_quark_effective_actions}. The effective action $\effAction$ is then defined after integrating out the spatial link variables~\cite{Langelage2011, Fromm2012, Langelage:2014vpa},
	\begin{align}
		Z &= \int{\dif{U_0}} e^{-\effAction}, \quad -\effAction := N_f\ln \left(\det Q_\text{S}\right) + \ln \int[\dif{U_{\upsilon}}] \exp(-S_G)   \left(\det Q_{\text{K}} \right)^{N_f}. \label{def:effectiveAction}
	\end{align}
	Gauge invariance of the original lattice action passes on to the effective action and constrains the remaining degrees of freedom to be traces of polynomials of temporal Wilson lines $W_{\mathbf{x}}:=\prod_{\tau =0}^{N_\tau-1} U_0(\mathbf{x}, \tau)$, i.e., $\effAction$ is a functional of $W_{\mathbf{x}}$ only, so $S^\text{eff}[U_0] = S^\text{eff}[W]$ \cite{Langelage2011}.  
	
	In order to enable analytic integration over the spatial links in the full effective action $\effAction$ (\ref{def:effectiveAction}), 
	combined strong-coupling and hopping parameter expansions are performed~\cite{Langelage2011, Langelage:2014vpa}. Note that successive truncated versions of $\effAction$ define a sequence of effective theories, 
	which are expected to converge to the full effective action within the radius of convergence of both expansions. 
	For this reason we speak about effective actions (plural) instead of effective action (singular) in the sequel.
	
	\subsection{The pure gauge effective actions}
	
	For the pure gauge contribution to the effective actions it is beneficial to resum the strong-coupling series by a character expansion and the linked cluster theorem. Additional resummations include a subset of corrections into the couplings of the effective action, and modify the functional form of the interaction~\cite{Langelage2011}. 	
	The pure gauge effective action $\pureGaugeEffAction$ with nearest-neighbour interactions only is given by \cite{Langelage2011}
	\begin{align}
		-\pureGaugeEffAction = \sum_{\langle \mathbf{x},\mathbf{y}\rangle} \text{ln}\left(1+\lambda_1 L_{\mathbf{x}}L_{\mathbf{y}}^*+\lambda_1 L_{\mathbf{y}}L_{\mathbf{x}}^*\right) + \mathcal{O}(\beta^{2N_\tau})\quad \text{with}\quad  L_{\mathbf{x}} := \tr W_{\mathbf{x}}. \label{eq:pureGaugeAction} %In "\mathcal{O}(\beta^{2N_\tau})" habe ich "u" durch "\beta" ersetzt, weil "u" zum jetzigen Zeitpunkt noch nicht eingeführt wurde.
	\end{align}
The effective coupling $\lambda_1 = \lambda_1(\beta,N_\tau)$ is a function of the original couplings of the full lattice theory. Its resummed form used within this work is shortly discussed in appendix \ref{sec:DiscussionLambda1Expression}.
In the pure gauge limit, $\kappa = 0$, we simply have $\effAction = \pureGaugeEffAction$. Higher order corrections of  the character expansion imply interactions over arbitrary distances as well as interactions between Polyakov loops in higher representations \cite{Langelage2011}.

	\subsection{The kinetic quark effective actions \label{sec:Introduction_kinetic_quark_effective_actions}}
	
	For the quark effective action the kinetic quark determinant $\det Q_{\text{K}}$ has to be expanded. This is done using the identity $\det(\cdot) = \exp\left(\tr\ln(\cdot)\right)$ followed by a Taylor-expansion of the logarithm and the exponential function. Afterwards, resummation techniques have been used to rewrite the expression as a quark effective action $\quarkEffAction$~\cite{Langelage:2014vpa}. It can be expressed in terms of rational functions of untraced temporal Wilson lines,
	\begin{align}
		W_{nm\bar n \bar m}(W_{\mathbf{x}}) &= \tr \left( \frac{(h_1 W_{\mathbf{x}})^m}{(1+h_1 W_{\mathbf{x}})^n}\frac{(\bar h_1 W_{\mathbf{x}}^{-1})^{\bar m}}{(1+\bar h_1 W_{\mathbf{x}}^{-1})^{\bar n}}\right)\;, \quad W^\pm_{nm\bar n \bar m} = W_{nm00}\pm W_{00\bar n \bar m}.  \label{fractionalWilsonLoop}
	\end{align}
For $N_c = 3$ the $W_{nm\bar n \bar m}$ can always be represented in terms of the Polyakov loops $L$ and $L^*$ by generating function techniques \cite{Glesaaen2016} or the Cayley-Hamilton theorem \cite{Rindlisbacher:2015pea}. A third variant (which in view of the authors is the easiest and least error prone method) is discussed in appendix~\ref{sec:Relation_EVs_Polyakov_loops}.
	
	The quark effective action with leading-order corrections is \cite{Langelage:2014vpa}
	\begin{align}
		-\quarkEffAction &= N_f\ln \left(\det Q_\text{S}\right)-2N_f h_2\sum_{\langle \mathbf{x},\mathbf{y}\rangle} W_{1111}^-(W_{\mathbf{x}})W_{1111}^-(W_{\mathbf{y}}) + \mathcal{O}(\kappa^4)\notag\\
		 h_2 &= \frac{N_\tau\kappa^2}{N_c}(1+\dots)\label{eq:LOKineticAction}.
	\end{align}
	In the strong-coupling limit, $\beta = 0$, one simply has $\effAction = \quarkEffAction$.	The kinetic quark effective actions have been derived to $\mathcal{O}(\kappa^4)$ \cite{Neuman2015} and in the cold and dense regime to $\mathcal{O}(\kappa^8)$ \cite{Glesaaen2016}. In general, a kinetic quark effective action of order $\mathcal{O}(\kappa^{2n})$ implies non-local interactions between lattice sites, that can be connected by closed spatial quark paths of a Manhattan distance of at most $2n$.
	
	\subsection{Mixed contributions}
	Away from the strong-coupling ($\beta=0$)  or heavy-quark ($\kappa=0$) limit, all effective couplings depend on 
	all original QCD parameters, as shown in the derivations in~\cite{Fromm2012, NeumanPhD}, extending the validity of the derived effective actions to finite $\beta$ and $\kappa$. The full effective action is then approximated by
	\begin{align}
		\effAction(\beta, \kappa, N_\tau) \approx \quarkEffAction(\beta, \kappa, N_\tau) + \pureGaugeEffAction(\beta, \kappa, N_\tau).
	\end{align}
In the present work we consider actions correct to $\mathcal{O}(u^n \kappa^m)$ with $n+m \le 4$ as specified in appendix \ref{sec:DetailsEffectiveActionUsed}.	
	
	\section{Mean-field approximations for the effective theories \label{sec:General_mean_field_discussion}}
		
The standard mean-field (smf) approximation corresponds to expanding the action in fluctuations of the field variables
around a self-consistently determined mean field.  In the context of the effective theories this corresponds to rewriting the Polyakov loops in terms of their mean-field values $l, \bar{l}$ and the fluctuations around them, 
\begin{align}
L_{\mathbf{x}} = l +\delta L_{\mathbf{x}}, \quad L_{\mathbf{x}}^* = \bar{l} +\delta L_{\mathbf{x}}^*.
\end{align} 
Afterwards the action is Taylor-expanded to linear order in the fluctuations $\delta L_{\mathbf{x}}$ and $\delta L_{\mathbf{x}}^*$. 
In order to treat the differently truncated effective theories in the same way, we work with a formally generalised expression for the path integral,
	\begin{align}
		Z = \int [\dif{W_{\mathbf{x}}}]  \exp\left(\sum_{\mathbf{x}}\ln\det Q_\text{stat}^\text{loc}(L_{\mathbf{x}},L_{\mathbf{x}}^* )+\sum_{\langle \mathbf{x},\mathbf{y}\rangle} I(L_{\mathbf{x}}, L_{\mathbf{x}}^*, L_{\mathbf{y}}, L_{\mathbf{y}}^*)\right).\label{eq:pgStatic}
	\end{align}
Examples for the interaction are \eqref{eq:pureGaugeAction} and \eqref{eq:LOKineticAction}. In this generalised form, future  
extensions to interactions over longer ranges are straightforward. 
	
		In the following subsections mean-field approximations will be applied in three variations leading to different mean-field actions - each capturing different amounts of local fluctuations of the original effective theory and introducing different levels of complexity for solving the single-site problems. 
	
		\subsection{Standard mean-field approximation\label{sec:smf}}
Expanding the interaction $I$ up to linear order in the fluctuations  
one rewrites the effective action into local interactions between the Polyakov loops and the mean fields,
	\begin{align}
	S^{\text{eff}} =& S^{\text{eff}}_{\text{smf}} + \mathcal{O}(\delta L^2) = \sum_{\mathbf{x}} s_{\text{smf}}^{\text{eff}}(l, \bar{l}, L_{\mathbf{x}}, L_{\mathbf{x}}^*) + \mathcal{O}(\delta L^2)  \notag\\
	-s_{\text{smf}}^{\text{eff}}(l, \bar{l}, L, L^*) =& \ln\det Q^{\text{loc}}_{\text{stat}} (L,L^*) + dI(l, \bar{l}, l, \bar{l}) +d (L - l) \frac{\partial}{\partial l}I(l, \bar{l}, l, \bar{l}) \notag\\
	&+ d (L^* - \bar{l}) \frac{\partial}{\partial \bar{l}}I(l, \bar{l}, l, \bar{l}),
	\end{align}
	with the number of spatial dimensions $d$. This factorises the partition function, i.e., 
	\begin{align}
	         Z &\approx  z_{\text{smf}}^{N_s} \;,    \nonumber \\
		z_{\text{smf}} &= \int\dif{W}\exp\left(-s_{\text{smf}}^{\text{eff}}(l, \bar{l}, L, L^*)\right) \;,\label{eq:smf_partitionfunction}
	\end{align}
	where $N_s$ is the total number of (spatial) lattice sites. The task of computing the partition function is reduced to solving the single-site integral in \eqref{eq:smf_partitionfunction} together with the two coupled self-consistency equations
	\begin{align}
		l = \langle L \rangle_{\text{smf}}(l,\bar{l}) \quad\text{and}\quad \bar{l} = \langle L^* \rangle_{\text{smf}}(l,\bar{l}).
		\label{eq:smf_sc}
	\end{align}
	From the set of solutions to \eqref{eq:smf_sc}, i.e., the set of self-consistent mean fields, the one with the largest $z_{\text{smf}}$ represents the best approximation to the true partition function \eqref{eq:pgStatic}.
	Equivalently to solving the self-consistency relations \eqref{eq:smf_sc} one can also determine the saddle points of the smf free energy density \cite{Zinn-Justin202},
	\begin{align}
		a^{4} f_{\text{smf}}(l, \bar{l}) = -\frac{1}{N_{\tau}} \ln z_{\text{smf}}.\label{eq:def:f-smf}
	\end{align}
	Finding the best approximation to the free energy density corresponds to determining the set of saddle points of \eqref{eq:def:f-smf}, of which the one with the lowest $f_{\text{smf}}$
	%, hereinafter referred to as the minimizing self-consistent mean field, 
	determines the physically relevant result.
	This approach does not require $S$ or $S_{\text{smf}}$ to be real valued, and hence is applicable also  in the presence of non-vanishing 
	baryo-chemical potentials.	
	Note that one may also expand the contribution coming from the static determinant in the same manner. Here this is not done because the static determinant already is a single-site interaction.  
	
	\subsection{Self-consistent mean fields and the sign problem \label{sec:Mfs_and_sign_problem}}
	The self-consistency relation \eqref{eq:smf_sc} amounts to solving a variational problem for the two in general complex-valued variables $l$ and $\bar{l}$. 
	%Consequently, to actually solve this equation it is beneficial to consider physical expectations on the behaviour of the expectation values $\langle L \rangle$ and $\langle L^* \rangle$. 
	In the $\mathrm{SU}(3)$ pure gauge theory, the Polyakov loop is well-known to possess three degenerate vacuum solutions, which are related
	by $Z(3)$ centre symmetry. 
	%At $\mu_B = 0$ charge conjugation symmetry is not explicitly broken, implying $\langle L \rangle = \langle L^* \rangle^*$. 
	 In QCD with finite-mass quarks the centre symmetry is explicitly broken to favour the vacuum with Polyakov loops along the real axis, i.e., one has $\langle L \rangle = \langle L^* \rangle$. In this case \eqref{eq:smf_sc} reduces to a single field problem with $l = \bar{l} \in \mathbb{R}$. 
	 When a non-vanishing baryo-chemical potential is switched on,  $\mu_B\not = 0$, charge conjugation symmetry is explicitly broken 
	 and one has $\langle L \rangle \not= \langle L^* \rangle$, implying also $l \not = \bar{l}$ 
	in general~\cite{Dumitru:2005ng}.  
	
	In order to avoid this complication,  it was proposed in \cite{Fukushima:2006uv} to  perform mean-field approximations for the phase-quenched effective action, which is symmetric under charge conjugation even for $\mu_B\not = 0$.  This implies $\langle L \rangle_{\text{pq}} = \langle L^* \rangle_{\text{pq}}$.   As in the case $\mu_B = 0$, only a single self-consistent equation is needed,
	\begin{align}
		l_{\mathrm{pq}} = \langle L \rangle_{\text{smf},\mathrm{pq}}(l_{\mathrm{pq}}).\label{eq:smf_pq_sc}
	\end{align}
	 In order to get back to the physical situation with  $\langle L \rangle \not= \langle L^* \rangle$, the resulting phase-quenched mean field
	is used as $l=\bar{l}=l_\mathrm{pq}$ on the right hand side of  \eqref{eq:smf_sc}, and the expectation values are then evaluated
	by reweighting. 
	In a later work 
	%on mean field approximations for effective Polyakov loop theories
	\cite{Rindlisbacher:2015pea} it was argued that this introduces no further approximation,
	since the reweighting step is exact once all field values are integrated over when computing the expectation value. 
	However, this does not reduce \eqref{eq:smf_sc} to \eqref{eq:smf_pq_sc}, and hence such a calculation of expectation values differs from a self-consistent variational solution of \eqref{eq:smf_sc}, so that in 
	general 
	\begin{align}
	\langle L \rangle_{\text{smf}}(l_{\mathrm{pq}},l_{\mathrm{pq}}) &\neq \langle L \rangle_{\text{smf}}(l,\bar{l}), \nonumber \\
	\langle L^* \rangle_{\text{smf}}(l_{\mathrm{pq}},l_{\mathrm{pq}}) &\neq \langle L^* \rangle_{\text{smf}}(l,\bar{l}) .
	\end{align}
		 Alternatively, one may start with the mean-field action corresponding to \eqref{eq:smf_partitionfunction} and decompose it into two terms, which are symmetric and anti-symmetric under charge conjugation, respectively.\footnote{Beware that under charge conjugation one not only exchanges $L\leftrightarrow L^*$ but also the two mean fields $l \leftrightarrow \bar{l}$. This follows directly from the self-consistency relation \eqref{eq:def:f-smf}.} Then one neglects fluctuations at $\mathcal{O}(\delta L)$ of the anti-symmetric part. The resulting  self-consistency equations are symmetric under charge conjugation and \eqref{eq:smf_sc} reduce to \eqref{eq:smf_pq_sc}. This, however, %also%
		  amounts to an additional approximation at $\mu_B \not = 0$, because of the neglected fluctuations at $\mathcal{O}(\delta L)$ of the anti-symmetric part under charge conjugation of the mean-field action. We complete this discussion in section \ref{subsec:eval_decon} by explicitly demonstrating that results obtained from \eqref{eq:smf_sc} and \eqref{eq:smf_pq_sc} are not identical. 
 
 %Additionally discussed is the the impact of this additional approximation on the accuracy of the mean field approach. This impact is not obvious, because one expects the sign-problem to have a qualitative effect on the phase diagram of LQCD, but the sign-problem within the effective theories is known to be significantly more mild compared to full LQCD. 
	
	\subsection{Resummed mean-field approximation \label{sec:rmf}}
	
 In this section we propose a different truncation scheme that allows us to resum a subset of fluctuations to all orders, which will lead 
 to a considerable quantitative improvement of the approximation.
	As in section \ref{sec:smf}, 
	%we start by Taylor-expanding the two-point interaction $I$ in the fluctuations after writing $L_{\mathbf{x}} = l +\delta L_{\mathbf{x}}$ and $L_{\mathbf{x}}^* = \bar{l} +\delta L_{\mathbf{x}}^*$. Afterwards, 
	terms describing non-local fluctuations, i.e., terms of $\mathcal{O}(\delta L_{\mathbf{x}}\delta L_{\mathbf{y}})$ with $\mathbf{x} \not = \mathbf{y}$, are neglected.  But in contrast to the standard approach, we now keep all powers of purely local fluctuations, i.e., all terms $\sim\delta L_{\mathbf{x}}^n\delta L_{\mathbf{x}}^{*,m}$. The resulting expression can be resummed, and the partition function factorises again,
	\begin{align}
	Z &= z_{\text{rmf}}^{N_s} + \mathcal{O}(\delta L_{\mathbf{x}}\delta L_{\mathbf{y}}), \quad \mbox{with}\quad \mathbf{x} \not = \mathbf{y}, \nonumber \\
		z_\text{rmf} &= e^{-d I(l,\bar{l},l,\bar{l})}\int\dif{W}\det Q_\text{stat}^\text{loc}(W) e^{2d I(L,L^*,l,\bar{l})}.
		\label{eq:ssr}
	\end{align}
	Finding the saddle points of the resummed mean-field (rmf) free energy density $f_{\text{rmf}}(l,\bar{l})$ corresponds to solving
	\begin{align}
		\begin{pmatrix}
			\frac{\partial}{\partial l}\\
			\frac{\partial}{\partial \bar{l}}
		\end{pmatrix} I(l,\bar{l},l,\bar{l}) = \frac{2}{z_{\text{rmf}}}\int\dif{W}\det Q_\text{stat}^\text{loc}(W) e^{2d I(L,L^*,l,\bar{l})}	\begin{pmatrix}
		\frac{\partial}{\partial l}\\
		\frac{\partial}{\partial \bar{l}}
	\end{pmatrix} I(L,L^*,l,\bar{l}).\label{eq:rmf_sc}
	\end{align}
	Note that this is no longer identical to the usual self-consistency relations, i.e.,
	\begin{align} 
	l \not= \langle L \rangle_{\text{rmf}},\quad \bar{l} \not= \langle L^* \rangle_{\text{rmf}}.
	\end{align}
	 Following the same variational reasoning as in section \ref{sec:smf}, 
	  namely that the optimal mean field should minimize the free energy density, 
	 %where solving the self-consistency relation is equivalent to finding the saddle-points of the mean free energy density 
	 implies that \eqref{eq:rmf_sc} plays the role of the self-consistency relation in the case of resummed local fluctuations.\footnote{This is in contrast to a discussion in \cite{Greensite:2012xv} where the authors argue that the usually expected self-consistency relations hold even if corrections to the standard mean-field approximation are included, i.e., we would have  $l = \langle L \rangle_{\text{rmf}}$ and $\bar{l} = \langle L^* \rangle_{\text{rmf}}$. 
	 	%In their work it is said that their equation (46) clearly implies $l = \langle L \rangle_{\text{rmf}}$ and $\bar{l} = \langle L^* \rangle_{\text{rmf}}$ (our analog to their equation (47)). 
	 	However, the interaction term "J" in their equation (46) also depends on the variational parameters, implying that $l = \langle L \rangle_{\text{rmf}}$ and $\bar{l} = \langle L^* \rangle_{\text{rmf}}$ does not hold in general 
	 	%if corrections to the standard mean-field approximation are included
	 	.} Indeed, after undoing the resummation in \eqref{eq:rmf_sc}, i.e., neglecting terms at $\mathcal{O}(\delta L^2)$, one recovers the standard self-consistency relation \eqref{eq:smf_sc}.
	In section \ref{sec:eval_pure_gauge} we will demonstrate that this resummation scheme results in a significant increase of accuracy for determining the critical couplings of
	the first-order deconfinement transition.
	
	\subsection{Saddle-point approximation in angular parametrisation \label{sec:CA}}
	
	In the two previous variants of mean-field approximations the computation of the full partition function \eqref{eq:pgStatic} was reduced to a variational problem that involved solving a single-site path integral. Tackling the latter necessitates the use of numerical integration methods, which tend to get computationally expensive 
	once the integrand in \eqref{eq:ssr} gets sharply peaked.
	Looking at examples for the effective kinetic quark couplings, equations \eqref{eq:FullExpression:h1}--\eqref{eq:FullExpression:h33}, one finds that this is always the case for 
	 $\kappa, \mu_B \neq 0$ when $N_\tau$ is sufficiently large. In terms of physical parameters this corresponds to the low-temperature and finite-density regime. For sharply peaked integrands, however, standard saddle-point approximations are expected to give reliable results. This leads us to consider a third variant of mean-field approximations, which we will refer to as the angular mean-field (amf) approximation.
	
	Again, we start with the full partition function of our effective theory at hand, but perform a change of variables, where the Polyakov loops become functions of the angles $\phi_1$ and $\phi_2$ \cite{Gross:1983ju},
	\begin{align}
		L &\rightarrow L(\phi_1,\phi_2) = e^{i\phi_1} +  e^{i\phi_2} + e^{-i(\phi_1 + \phi_2)}\quad\text{and}\quad	L^* \rightarrow L(-\phi_1,-\phi_2). \label{eq:ChangeOfVariables}
	\end{align}
	This induces a Jacobian which may be shifted into an effective potential $V_{\text{eff}}$ \cite{Gross:1983ju},
	\begin{align}
		&Z = \int\left[\dif{W_{\mathbf{x}}}\right]e^{-S^{\text{eff}}} = \frac{1}{(2\pi)^{2N_s}6^{N_s}}\int\left[\dif{\phi_{1,\mathbf{x}}}\right]\left[\dif{\phi_{2,\mathbf{x}}}\right]e^{-S^{\text{eff}} + V_{\text{eff}}}\label{eq:Change_to_reduced_Haar_measure}\\
		&\text{with } V_{\text{eff}} = \sum_{\mathbf{x}} V_{\mathbf{x}} = \sum_{\mathbf{x}}\ln\left(27 - 18\left|L_{\mathbf{x}}\right|^2 + 8\ReN\left(L_{\mathbf{x}}^3\right) - \left|L_{\mathbf{x}}\right|^4\right)\label{eq:Effective_Potential}.
	\end{align}
	Until now no approximations were performed (except for the truncations involved in the derivation of the effective action). A standard leading-order saddle-point approximation is performed for the angles $\phi_1$ and $\phi_2$ around the saddle points $\Phi_1$ and $\Phi_2$ of $\hat{S}^{\text{eff}} := S^{\text{eff}} - V_{\text{eff}}$,
	\begin{align}
		Z =& \frac{1}{(2\pi)^{2{N_s}}6^{N_s}}\int\left[\dif{\phi_{1,\mathbf{x}}}\right]\left[\dif{\phi_{2,\mathbf{x}}}\right]\exp\left(
		\vphantom{\left.\frac{\partial}{\partial L_{\mathbf{x}}}\hat{S}^{\text{eff}}[L,L^*]\right\vert_{\substack{L = L(\Phi_1,\Phi_2)}}}
		-\hat{S}^{\text{eff}}[L(\Phi_1,\Phi_2), L(-\Phi_1,-\Phi_2)]\right.\notag\\
		&\left.- \sum_{\mathbf{x}} \left.\frac{\partial}{\partial L_{\mathbf{x}}}\hat{S}^{\text{eff}}[L,L^*]\right\vert_{\substack{L = L(\Phi_1,\Phi_2) \\ L^* = L(-\Phi_1,-\Phi_2)}} \sum_{i=1,2}\frac{\partial L(\Phi_1,\Phi_2)}{\partial \Phi_i} \delta \phi_{i,\mathbf{x}} \right. \notag\\
		&\left.- \sum_{\mathbf{x}} \left.\frac{\partial}{\partial L^*_{\mathbf{x}}}\hat{S}^{\text{eff}}[L,L^*]\right\vert_{\substack{L = L(\Phi_1,\Phi_2) \\ L^* = L(-\Phi_1,-\Phi_2)}} \sum_{i=1,2}\frac{\partial L(-\Phi_1,-\Phi_2)}{\partial \Phi_i} \delta \phi_{i,\mathbf{x}} + \mathcal{O}(\delta\phi^2) \right). \label{eq:Partition_Function_SaddlePointApproximation}
	\end{align}
	The saddle-point equations for the $\Phi_1$ and $\Phi_2$ are (written in the form of a Jacobian matrix multiplied with a vector)
	\begin{align}
		0 =& \begin{pmatrix}
			\frac{\partial L(\Phi_1,\Phi_2)}{\partial \Phi_1} & \frac{\partial L(-\Phi_1,-\Phi_2)}{\partial \Phi_1} \\
			\frac{\partial L(\Phi_1,\Phi_2)}{\partial \Phi_2} & \frac{\partial L(-\Phi_1,-\Phi_2)}{\partial \Phi_2}
		\end{pmatrix} \begin{pmatrix}
		\sum_{\mathbf{x}} \left.\frac{\partial}{\partial L_{\mathbf{x}}}\hat{S}^{\text{eff}}[L,L^*]\right\vert_{\substack{L = L(\Phi_1,\Phi_2) \\ L^* = L(-\Phi_1,-\Phi_2)}} \\
		\sum_{\mathbf{x}} \left.\frac{\partial}{\partial L^*_{\mathbf{x}}}\hat{S}^{\text{eff}}[L,L^*]\right\vert_{\substack{L = L(\Phi_1,\Phi_2) \\ L^* = L(-\Phi_1,-\Phi_2)}}
		\end{pmatrix}\label{eq:CA:SaddlePoint}
	\end{align}
	and dictate that the second and third line of \eqref{eq:Partition_Function_SaddlePointApproximation} vanish. Thus, within our approximation the integrand of \eqref{eq:Partition_Function_SaddlePointApproximation} is independent on the field variables, $\phi_1$ and $\phi_2$, allowing us to easily perform the integration analytically. Afterwards, the partition function is 
	$Z \approx z_{\text{amf}}(\Phi_1,\Phi_2)^{N_s}$, where
	\begin{align}
		z_{\text{amf}}(\Phi_1,\Phi_2) :=& \frac{1}{6}\exp\left(-\frac{1}{N_s}\hat{S}^{\text{eff}}[L(\Phi_1,\Phi_2), L(-\Phi_1,-\Phi_2)]\right),\label{eq:def:CA_SingleSitePartitionFunction}
	\end{align}
	and the free energy density is approximated by $a^4 f\approx a^4f_{\text{amf}}$, with
	\begin{align}
		a^4f_{\text{amf}} := -\frac{1}{N_\tau} \ln z_{\text{amf}}(\Phi_1,\Phi_2) = \frac{1}{N_\tau N_s}\hat{S}^{\text{eff}}[L(\Phi_1,\Phi_2), L(-\Phi_1,-\Phi_2)] - \frac{\ln(6)}{N_\tau},
	\end{align}
	i.e. the free energy density is simply determined via the action density. 
	
	No single-site integration has to be performed any more for this evaluation of $Z$. The only task left is to determine the saddle points \eqref{eq:CA:SaddlePoint}, where all physically relevant information is contained in the derivatives of $\hat{S}^{\text{eff}}$ with respect to the Polyakov loops. We therefore multiply \eqref{eq:CA:SaddlePoint} with the inverse of the Jacobian appearing in \eqref{eq:CA:SaddlePoint}. After setting $l := L(\Phi_1,\Phi_2)$ and $\bar{l} := L(-\Phi_1,-\Phi_2)$ the relevant saddle-point equations read
		\begin{align}
		\sum_{\mathbf{x}}\left.\frac{\partial}{\partial L_{\mathbf{x}}}\hat{S}^{\text{eff}}[L,L^*]\right\vert_{\substack{L = l \\ L^* = \bar{l}}} = 0 \quad \text{and} \quad \sum_{\mathbf{x}}\left.\frac{\partial}{\partial L^*_{\mathbf{x}}}\hat{S}^{\text{eff}}[L,L^*]\right\vert_{\substack{L = l \\ L^* = \bar{l}}} = 0,
	\end{align}
	which are equivalent to determining the saddle points of $f_{\text{amf}}$ as a function of $l$ and $\bar{l}$,
	\begin{align}
		\frac{\partial}{\partial l}a^4f_{\text{amf}}(l, \bar{l}) = 0\quad\text{and}\quad \frac{\partial}{\partial \bar{l}}a^4f_{\text{amf}}(l, \bar{l}) = 0.
	\end{align}
	Following the same steps for the expectation values of the Polyakov loops one finds $\langle L \rangle \approx \langle L \rangle_{\text{amf}}$ and $\langle L^* \rangle \approx \langle L^* \rangle_{\text{amf}}$, where
	\begin{align}
		\langle L \rangle_{\text{amf}} = l\quad\text{and}\quad\langle L^* \rangle_{\text{amf}} = \bar{l},
	\end{align}
	which are just the expected self-consistency relations for two mean fields $l$ and $\bar{l}$. The Polyakov loops $l, \bar{l}$ which are the stationary points of the action determine the self-consistent mean fields. 
	%This justifies naming this approach the classical approximation (amf). However, even if one had access to the exact effective action, the amf approximation of the effective theory does not correspond to the classical approximation of LQCD, because the spatial link integrals have been integrated out and, thus, part of the path integral has been solved exactly.
	
	In contrast to the previous two mean-field approaches, due to the absence of a (single-site) path integral a reweighting procedure is not well defined. Thus, determining the self-consistent mean fields with respect to the phase-quenched ensemble always leads to $\langle L \rangle_{\text{amf}} = \langle L^* \rangle_{\text{amf}}$, even at $\mu_B\not = 0$, and is therefore not applicable within this approach.
	
	Another observation can be made in the case of the trivial effective theory, i.e., the one obtained by setting all couplings to $0$. Then, the integrals in \eqref{eq:pgStatic} simply result in the volume of the gauge group. Because of the normalization of the Haar measure one finds $Z = 1$. Using the amf approximation one finds for the trivial effective theory the self-consistent mean fields at $l = 0 = \bar{l}$. The single-site partition function, however, gives $z_{\text{amf}} = 27/6 = 9/2$ resulting in $Z = (9/2)^{N_s} + \mathcal{O}(\delta \phi^2)$. Thus, the normalization of the Haar measure is lost within this approximation scheme. This will be important for applications in section \ref{sec:sys}.
	
	\section{Mean-field evaluation of the effective theories \label{sec:Evaluation}}
	
	We now turn to the application of the mean-field approximations derived in the previous section. 
	In order to express our results in physical units we set a scale by expressing the lattice spacing in terms of the well-known $r_0$ parameter in the pure gauge limit \cite{Necco:2001xg},
	\begin{align}
		a(\beta) =& r_0\exp\left(-1.6804 - 1.7331(\beta-6)+0.7849(\beta-6)^2 - 0.4428(\beta-6)^3\right)\notag\\
		&\text{for } 5.7 \le \beta \le 6.92. \label{eq:lattice_spacing}
	\end{align}
	This should be a good approximation also for our case, where only heavy quarks are considered.
	Further we use hopping-resummed expressions for the 
	%pion mass $am_\pi$ and 
	baryon mass $am_B$ in lattice units  \cite{Smit_2002} 
	with their leading gauge corrections,
	\begin{align}
		%am_\pi(\beta, \kappa) &= \mathrm{arccosh}\left[1 + \frac{(M^2 - 4)(M^2 - 1)}{2M^2-3}\right] - 24\kappa^2\frac{u}{1-u} + \dots\\
		am_B(\beta, \kappa) &= \ln\left[\frac{M^3(M^3-2)}{M^3-\frac{5}{4}}\right] - 18\kappa^2\frac{u}{1-u} + \dots,
	\end{align}
	where $M\equiv 1/(2\kappa)$. For a given $N_\tau$ we then relate the bare lattice parameters to the temperature $T$ and 
	%pion 
	baryon mass $m_B$ in the continuum by
	\begin{align}
		am_B(\beta, \kappa)/a(\beta) &= m_B\\
		N_\tau am_B(\beta, \kappa) &= m_B/T.
	\end{align}
	
	At $\mu_B = 0$ the self-consistent mean fields are determined via a search for the local minima of the corresponding mean-field free energy density $f_{\text{mf}}(l,l)$, using the constraint $l = \bar{l}$. 
	At $\mu_B \not = 0$ we allow $l \not = \bar{l}$ and determine the local minima, maxima and saddle points of $f_{\text{mf}}(l,\bar{l})$.  
	Thus, if not explicitly stated otherwise, self-consistent mean fields are determined without resorting to the phase-quenched approach.
	
	\subsection{Deconfinement transition in the pure gauge theory \label{sec:eval_pure_gauge}}
	
	\begin{figure}[!t]
		
		\centering
		\includegraphics[height=5cm]{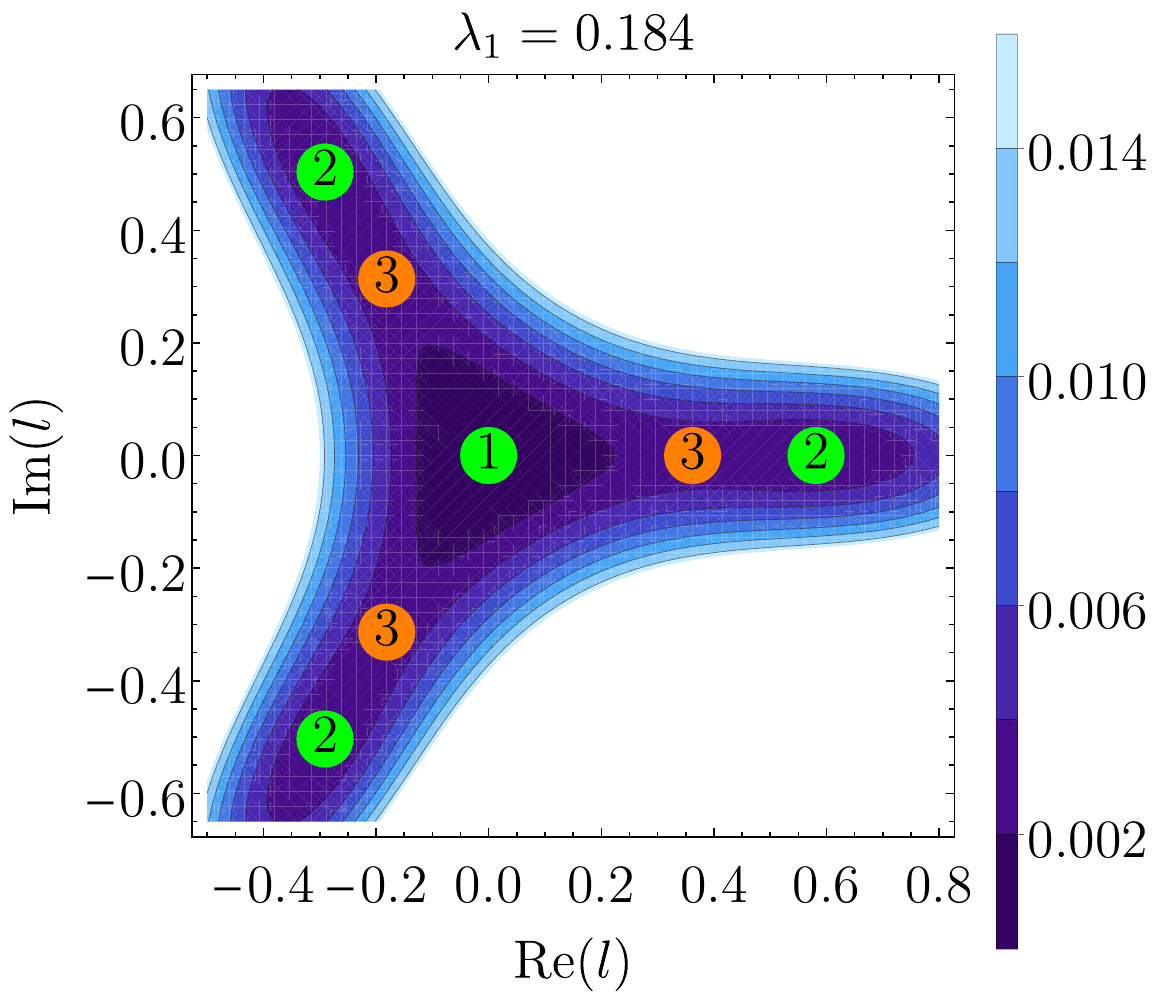}\hspace*{1cm}
		\includegraphics[height=5.2cm]{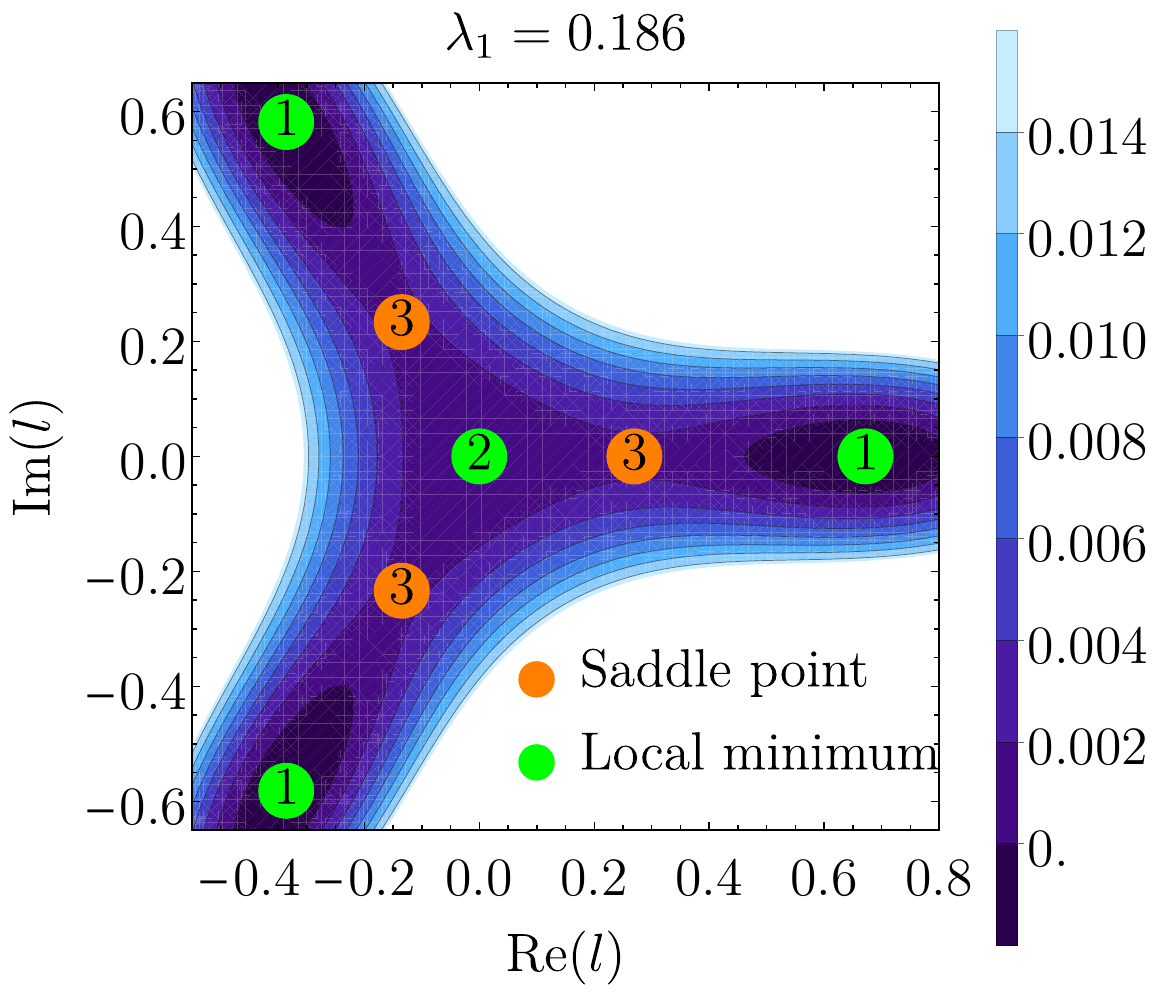}
		\includegraphics[height=5.2cm]{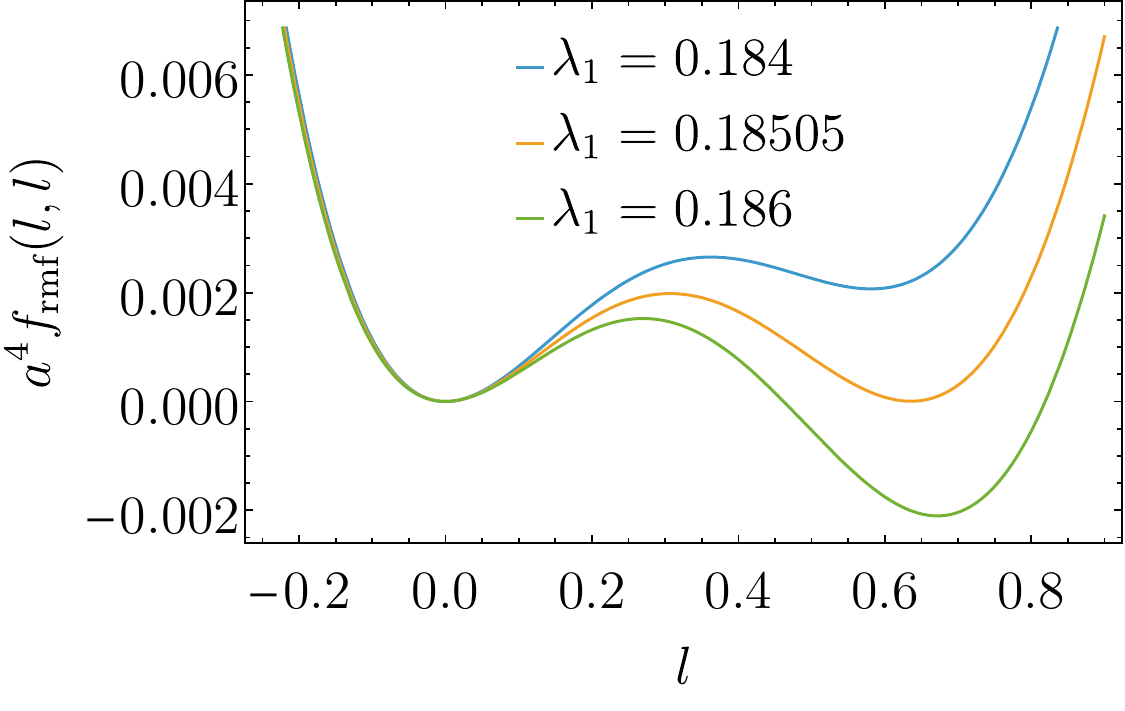}
		\caption{Top row: the resummed mean-field free energy density $a^4f_{\text{rmf}}$ in the complex mean field plane with $\bar{l} = l^*$. Bottom row: $a^4f_{\text{rmf}}$ for identical real-valued mean fields, $l = \bar{l}\in \mathbb{R}$.}
		\label{fig:pure_gauge_fRmf}
		
	\end{figure}
	We begin with the simplest effective theory representing  the four-dimensional pure gauge theory, (\ref{eq:pureGaugeAction}), truncated to only 
	one effective coupling for nearest-neighbour interactions, $\lambda_1$. The general procedure is illustrated at the example of
	 the resummed mean-field approach. 
	 	Considering the complex mean-field plane with $\bar{l} = l^* \in \mathbb{C}$ leads to the two contour plots of $f_{\text{rmf}}(l,\bar{l})$ at coupling values slightly below (upper left of figure \ref{fig:pure_gauge_fRmf}) and above  (upper right of figure \ref{fig:pure_gauge_fRmf}) the symmetry breaking
		transition. The $Z(3)$ center symmetry of the original action is also obvious in $f_{\text{rmf}}(l,\bar{l})$. 
		The minimizing self-consistent mean fields correspond to the local minima of $f_{\text{rmf}}(l,\bar{l})$, labelled as "1" in  
		figure~\ref{fig:pure_gauge_fRmf}.  Since for the larger coupling the three symmetry-broken minima are degenerate,  it is sufficient to pick
		the one with a real field and work with the constraint $\bar{l} = l \in \mathbb{R}$. This results in the 
		 the free energy density $a^4 f_{\text{rmf}}$ in figure~\ref{fig:pure_gauge_fRmf} (lower plot) around the deconfinement transition, which 
		 happens at the critical coupling at which the minima are degenerate, and it is of first order.

\begin{figure}[!t]
	 	
	 	\centering
	 	
	 	\includegraphics[width=1.0\linewidth]{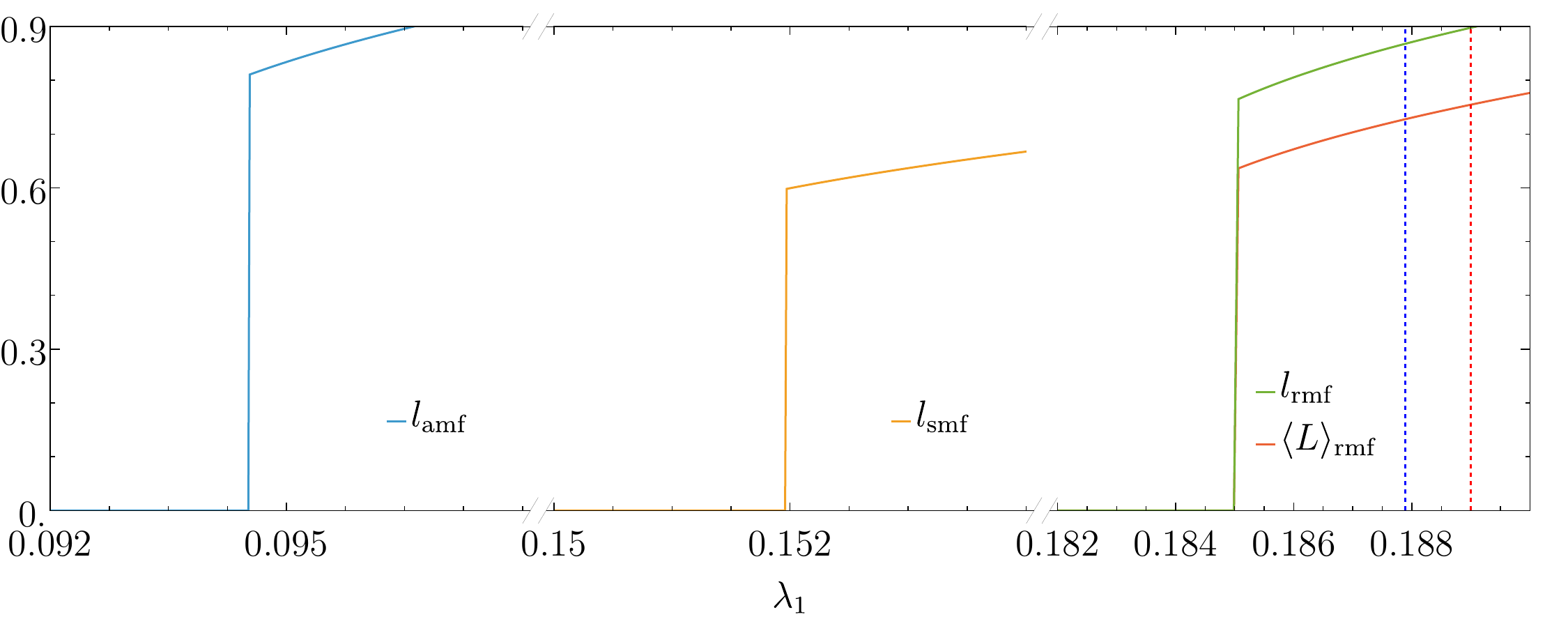}
	 	\caption{The minimizing self-consistent mean field in the amf approach (blue solid line), smf approach (orange line) and rmf approach (green line) versus the pure gauge effective coupling $\lambda_1$. Also shown is the expectation value of the Polyakov loop (red solid line) obtained via rmf. Vertical lines signal the critical $\lambda_1$ as obtained from series expansion techniques \cite{Kim:2019ykj} (red dashed line) and Monte Carlo simulations~\cite{Langelage2011} (blue dashed line).}
	 	\label{fig:resummedUnresummed}		
	 \end{figure}

	We now compare the predictive power of the different mean-field approaches.
	Figure~\ref{fig:resummedUnresummed} shows the minimizing self-consistent mean field as a function of the effective coupling for
	the three approaches discussed in  section~\ref{sec:General_mean_field_discussion}.
	 All of them show the expected jump from zero
	in the symmetric phase to a non-zero value in the broken phase at some critical coupling $\lambda_{1,c}$.
	 For comparison, the dashed red line marks the transition  obtained by high-order series expansion techniques~\cite{Kim:2019ykj}, whereas the dashed blue line highlights the transition obtained by simulations \cite{Langelage2011} of the same effective theory. 
	 
	The large variation in the mean field critical couplings demonstrates the influence of incorporating local fluctuations to a different extent. The prediction of the amf approximation, which contains no local fluctuations beyond leading order, is $\lambda_{1,c} \approx 0.09$  with $\approx 49$\% relative error compared to the simulation prediction. The smf approach, which includes local fluctuations beyond leading order coming from the Haar measure, shows an improved $\lambda_{1,c} \approx 0.15$  with $\approx 19$\% relative error. Finally the rmf approach, which includes all local fluctuations, 
	expectedly gives the best result of the three approaches with  $\lambda_{1,c} \approx 0.18505$ at only $\approx 1.5$\% relative error. As discussed in section~\ref{sec:rmf}, the minimizing self-consistent mean field and the expectation value of the Polyakov loop split up within this non-linear resummation
and are shown separately in figure~\ref{fig:resummedUnresummed}. Interestingly, the difference only affects the expectation value, but not the 
location of the transition.

\begin{figure}[!t]
		
		\centering
		
		\includegraphics[width=0.5\linewidth]{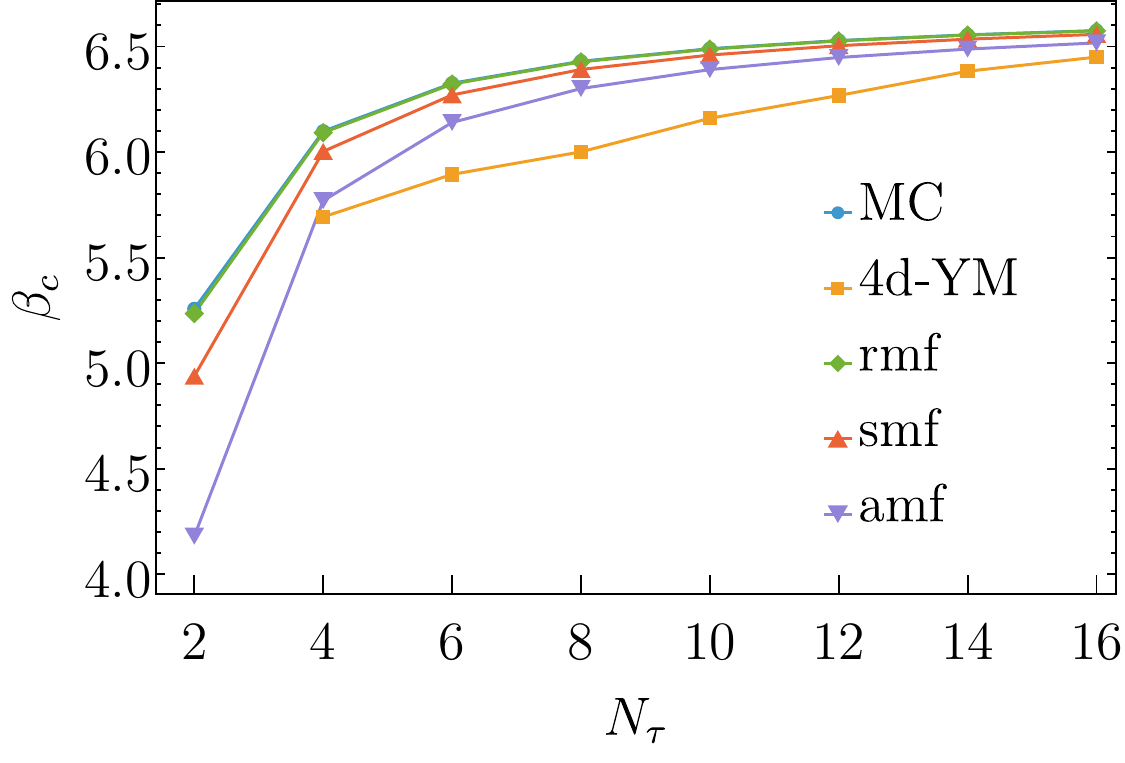}
		\caption{Comparison of the critical $\beta_c$ as obtained from the one-coupling effective theory \eqref{eq:pureGaugeAction} using the three mean-field approaches, as well as simulation results of the same effective theory (MC) \cite{Langelage2011} and of 4 dimensional Yang-Mills theory (4D-YM)~\cite{Kogut:1982rt,Fingberg:1992ju}. }
		\label{fig:betaC}		
	\end{figure}

Finally, the critical coupling $\lambda_{1,c}$ can be converted to a prediction of critical LQCD couplings by inverting (\ref{eq:Lambda1Expression}).
The resulting critical lattice coupling as a function of $N_\tau$ is shown in figure \ref{fig:betaC} for all three mean-field approaches, and compared
to a Monte Carlo evaluation of the effective theory as well as of the 4d Yang-Mills theory. Some care is in order when interpreting this figure: While
the resummed mean-field result is indistinguishable from the Monte Carlo solution of the effective theory, the other approaches are closer to the full 4d 
Yang-Mills result. This is due to an apparent partial cancellation of different systematic errors: truncation of the series in the derivation of the effective 
theories on one hand, and mean-field treatment of the latter on the other.  This can therefore not be taken as
a quality measure for the mean-field approaches, which can only be judged within the effective theory.
	
	\subsection{Deconfinement transition for QCD with heavy quarks  \label{subsec:eval_decon}}
	
	\begin{figure}[!t]\centering
				\begin{subfigure}[t]{.48\textwidth}
			\centering
			\includegraphics[width=\linewidth,page=1]{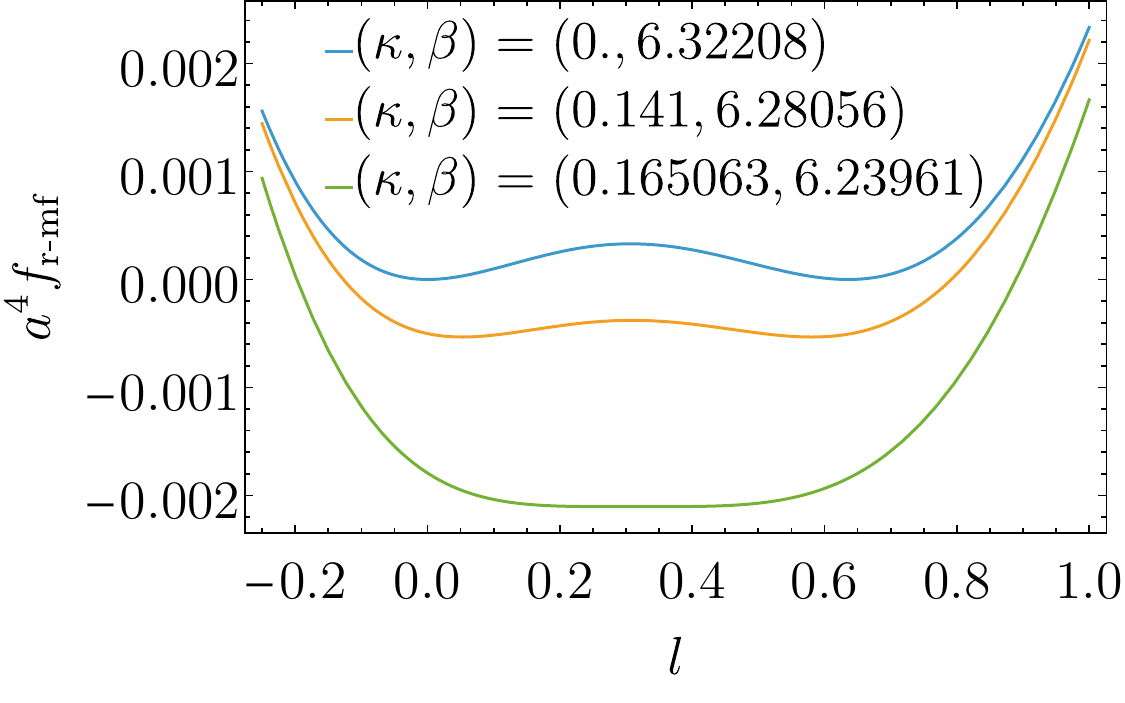}
			\caption{The free energy density versus the mean field $l$ for $N_\tau = 6$ at the deconfinement transition in the pure gauge limit (blue line), at intermediate values of $\kappa$ (orange line) and at the critical end-point (green line).}
			\label{fig:fMf}
		\end{subfigure}
		\quad
		\begin{subfigure}[t]{.48\textwidth}
			\centering
		\includegraphics[width=\linewidth,page=1]{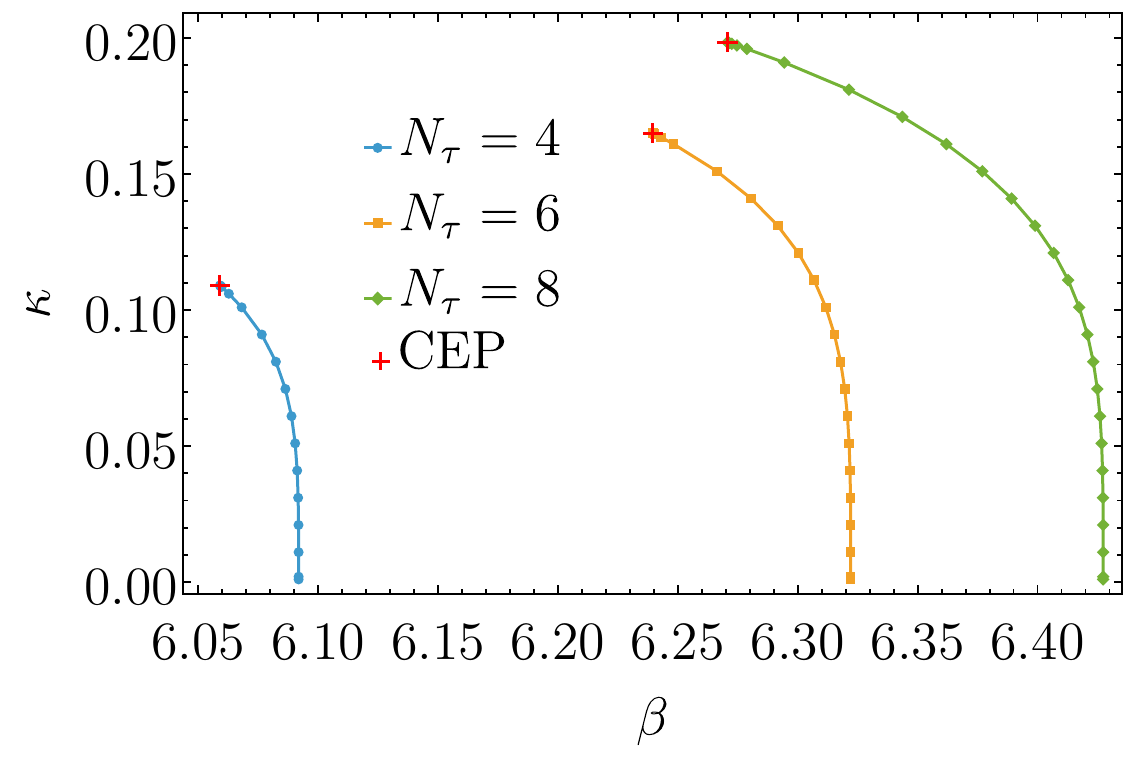}
			\caption{First-order deconfinement transition lines for $N_\tau = 4,6,8$ in the $\beta$-$\kappa$ plane ending in a second order critical end-point (CEP) (red cross).  Lines are a guide to the eye.}
			\label{fig:deconLine}	
		\end{subfigure}%
		\caption{The free energy density and first-order deconfinement transition lines obtained within the resummed mean-field approach for $N_f = 2$ and $\mu_B = 0$.}
		\label{fig:decon}
	\end{figure}
	
	We now consider an extended effective theory that approximates LQCD with heavy but  finite-mass quarks, i.e., we also include the static
	determinant as well as the kinetic quark action (\ref{eq:LOKineticAction}), with details as specified in appendix \ref{sec:qeff}.	
	Finite-mass quarks explicitly break the center symmetry of QCD, which is correctly reflected in the effective theory. As long as $\mu_B=0$,
	this explicit breaking favours real Polyakov loops and thus we continue to keep real valued and identical mean fields, $\bar{l} = l \in \mathbb{R}$. 
	 For the effective theory to describe QCD, the effective couplings $\{\lambda_1, h_1, h_2\}$ are completely fixed by the choice of QCD
	parameters $\{\beta,\kappa,N_f, N_\tau\}$, in terms of which we will express our following results.
	
	For $\mu_B = 0$, $N_f = 2$ and $N_\tau = 6$ we now repeat the previous analysis, slowly increasing $\kappa$ starting from the pure gauge 
	limit $\kappa=0$, and determining the corresponding critical coupling $\beta_c$ by the degeneracy of the symmetric and symmetry-broken
	minima in the free energy density, as shown in figure \ref{fig:fMf}. 
         As expected, with increasing $\kappa$ (decreasing quark mass) the barrier height in the mean-field free energy reduces, thus weakening the first-order transition until it disappears at a critical end point (green line) at $(\kappa_c, \beta_c) \approx (0.165063, 6.23961)$. The value of $\kappa_c$ deviates by $36\%$ from that of effective theory simulations~\cite{Pham:2021ftz}. This comparatively larger error is expected for the location of a critical end point, because the dynamics in its
         neighbourhood is driven by long-range fluctuations, causing mean-field treatments to break down. For completeness we note that
         the deviation of $\kappa_c$  from the simulation results of full LQCD \cite{Cuteri:2020yke} amounts to nearly $90\%$ in $\kappa$, 
         i.e. the two kinds of systematic errors add up
         in this case. This is easy to understand since long-range couplings are also missing in the derivation of the effective theory. 
         
         The resulting phase diagram for $N_\tau = 4,6,8$ is shown in figure \ref{fig:deconLine}. Each first-order line comes with a respective critical end point. Both $\kappa_c, \beta_c$ increase monotonously with $N_\tau$, but $\kappa_c$ exceeds the expected continuum limit value of $1/8$ for $N_\tau \ge 6$. This behaviour is consistent with earlier simulations of a similar effective theory \cite{Fromm2012} and can be explained by the truncation of the hopping expansion, see also section \ref{sec:sys}.

\begin{figure}[!t]
		
		\centering
		\begin{subfigure}[t]{.48\textwidth}
			\centering
			\includegraphics[width=\linewidth,page=1]{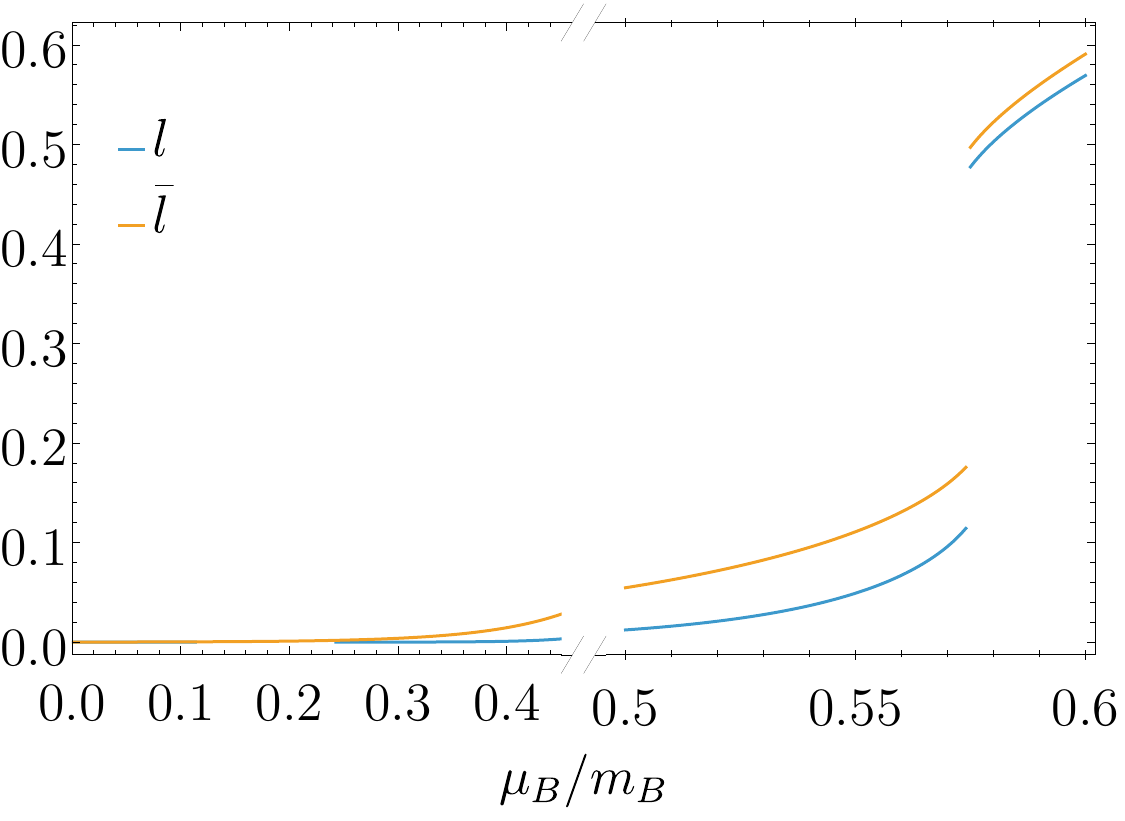}
			\caption{The minimizing self-consistent mean fields $l$ (orange line) and $\bar{l}$ (blue line) for varying $\mu_B/m_B$ in the vicinity of the first-order deconfinement transition.}
			\label{fig:mfsFiniteMuB}
		\end{subfigure}%
		\quad
		\begin{subfigure}[t]{.48\textwidth}
			\centering
		\includegraphics[width=\linewidth]{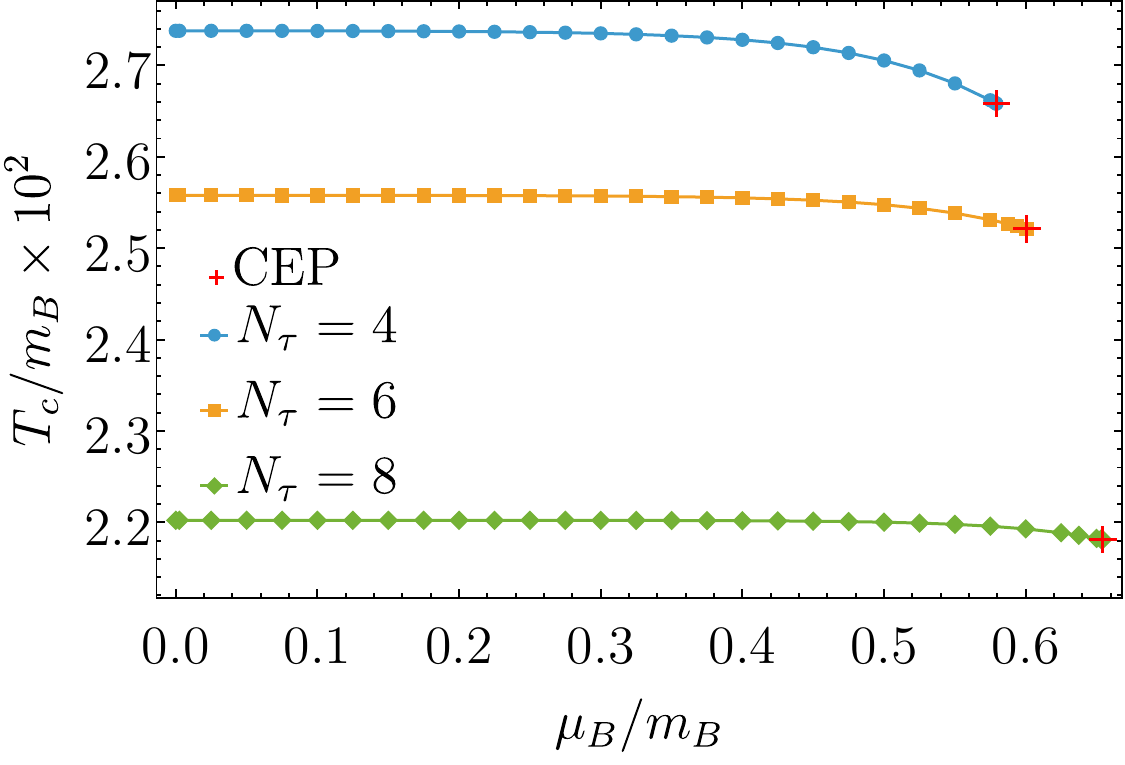}
		\caption{The first-order lines of the deconfinement transition in the $\mu_B/m_B$--$T/m_B$ plane as well as their critical end points (red cross). Lines are added to guide the eye.}
		\label{fig:decon_finiteMuB_phaseDiagram}
	\end{subfigure}
	\caption{The minimizing self-consistent mean fields around the first-order deconfinement transition at $\mu_B \not = 0$ for $N_\tau = 6, N_f = 2$ with $m_B = 22.5\text{ GeV}$ and $T/m_B = 0.025$ as well as the associated phase diagrams for $N_\tau = 4, 6$  and $8$. Data points have been obtained using the rmf approach without resorting to the phase-quenched approximation}
	\label{fig:mfsFiniteMuB:Both}
	\end{figure}

	Turning to $\mu_B \not= 0$,  charge conjugation symmetry is explicitly broken and we must consider the more general case 
	of unequal minimising mean fields, i.e., we have $l \not = \bar{l}$.
	 The mean fields obtained within the rmf approach are shown in figure \ref{fig:mfsFiniteMuB} for $N_f = 2$, $N_\tau = 6$, $\kappa \approx 0.055$ and $\beta \approx 6.313$ for varying $\mu_B/m_B$. These bare parameters correspond to $m_B = 22.5\text{ GeV}$ and $T/m_B = 0.025$.  At a critical $\mu_B/m_B \approx 0.574$ the first-order deconfinement transition occurs and induces a discontinuous jump of $l, \bar{l}$. Note that the difference of the two mean fields $ l - \bar{l}$ is small in the centre symmetric phase and even reduces in the symmetry-broken phase.
	
	In figure \ref{fig:decon_finiteMuB_phaseDiagram} the critical temperature is shown for $N_f = 2$ and $N_\tau = 4, 6$  and $8$ as a function of $\mu_B/m_B$. For each $N_\tau$ the deconfinement transition is a first-order transition that ends in a critical end point at some $\mu_B/m_B > 0.55$.

\begin{figure}[!t]\centering
		\begin{subfigure}[t]{.48\textwidth}
			\centering
			\includegraphics[width=\linewidth,page=1]{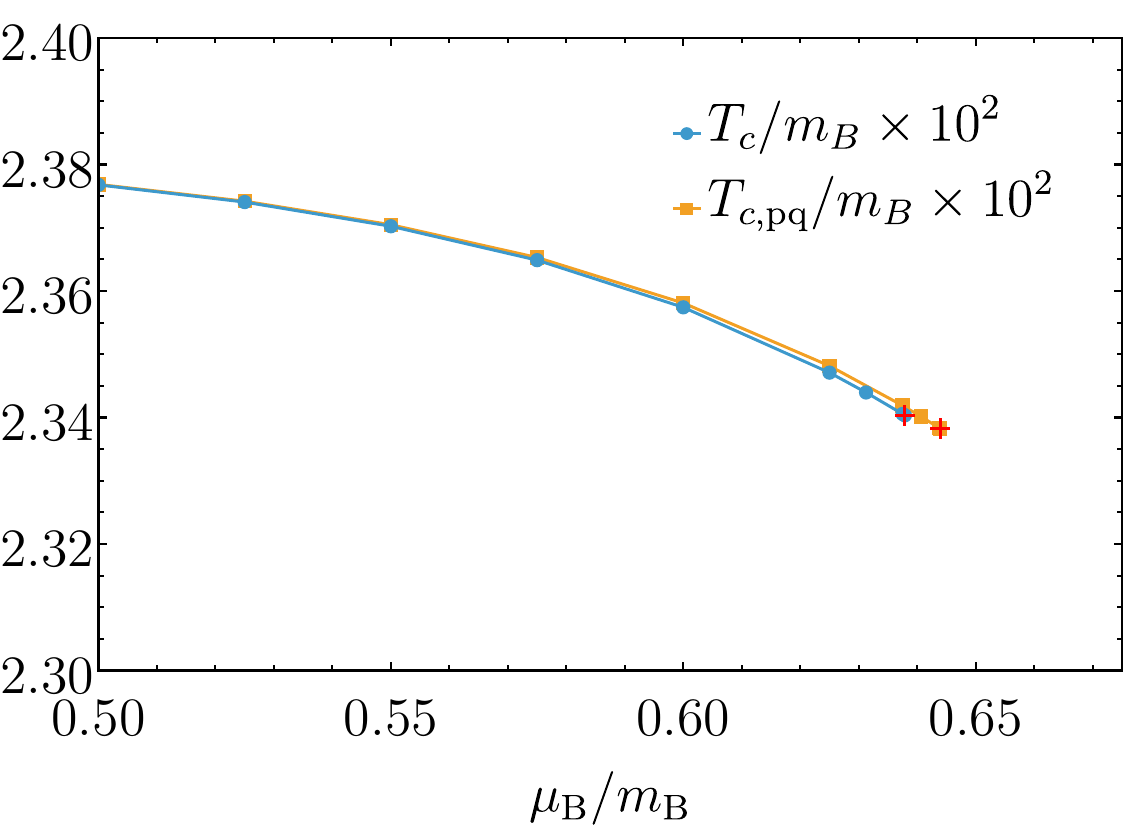}
			\caption{Comparison for the smf approach}
			\label{fig:S_MF_PQ_NOPQ_Comparison}
		\end{subfigure}%
		\quad
		\begin{subfigure}[t]{.48\textwidth}
			\centering
			\includegraphics[width=\linewidth,page=1]{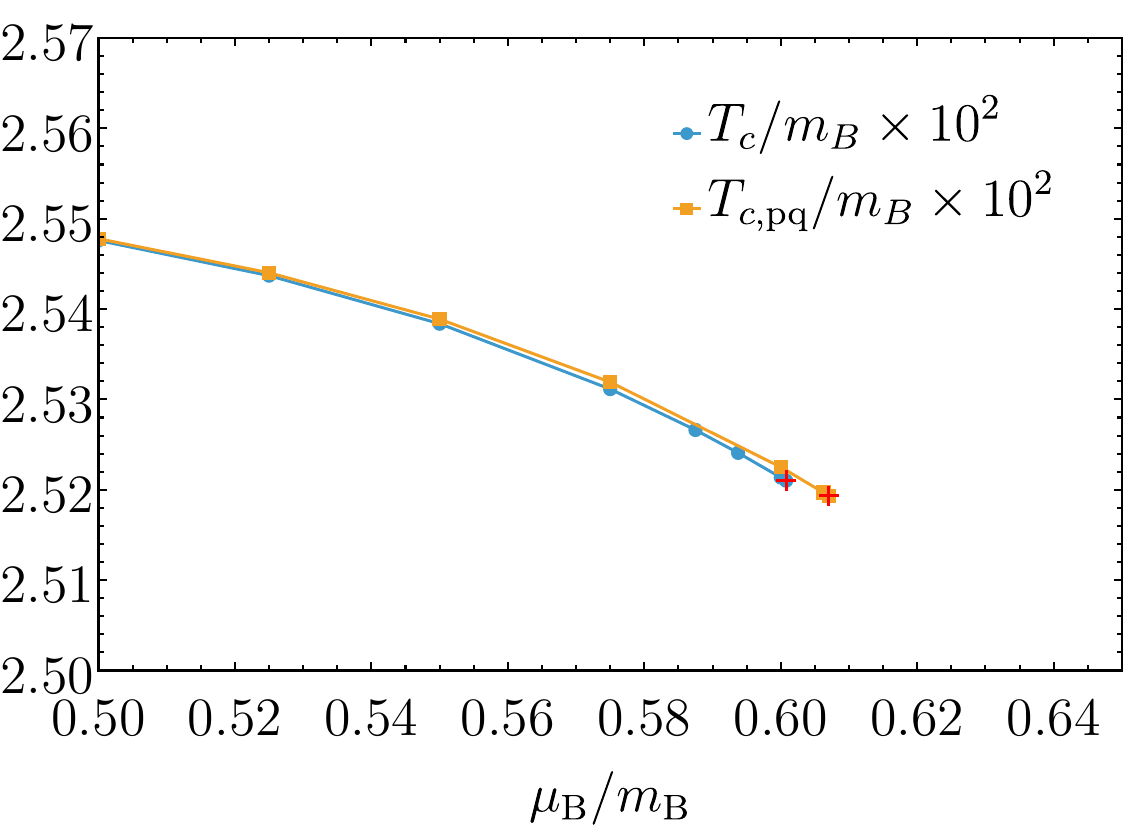}
			\caption{Comparison for the rmf approach}
			\label{fig:R_MF_PQ_NOPQ_Comparison}
		\end{subfigure}
		\caption{Comparison of the first-order deconfinement transition line with (blue dots) and without (orange dots) resorting to the phase-quenched approximation in the smf (left) and rmf (right) approach. Parameters are $N_f = 2, N_\tau = 6$ and $m_B = 22.5\text{GeV}$. Lines are added as a guide to the eye.}
		\label{fig:PQ_NOPQ_Comparison}
	\end{figure}

	We now compare the critical deconfinement line at $\mu_B\not = 0$ for the smf and rmf approaches, respectively, with and without resorting to the phase-quenched approximation, for $N_\tau = 6$. All other parameters are the same as in figure \ref{fig:mfsFiniteMuB:Both}. The comparisons are shown in figure \ref{fig:PQ_NOPQ_Comparison}. For both mean-field approaches the phase-quenched and non-phase-quenched first-order lines behave in a very similar way. Up to moderate values of $\mu_B$ the first-order lines obtained with and without phase quenching are approximately on top of each other. This is expected, because at $\mu_B = 0$ resorting to the phase-quenched ensemble is exact. When $\mu_B$ is further increased the first-order lines start to deviate from each other and end in critical end points at different locations of the phase diagram. This is in agreement  with our discussion in \ref{sec:smf}, where resorting to the phase-quenched ensemble was identified as an additional approximation at $\mu_B \not = 0$. However, the deviations of the critical end points are insignificant compared to the well-known large error bars of mean field for critical end points. The phase-quenched approximation therefore works very well within this parameter regime.
	
	\subsection{Low temperature and finite density\label{sec:Evaluation:LowTemperature}}
	
	The second region of interest of the QCD phase diagram is the low-temperature and finite-density regime. Numerical results based on complex Langevin or Monte Carlo simulations using the effective theories have been obtained several years ago \cite{Langelage:2014vpa, Glesaaen2016}, and we will compare them to our mean-field results. 
	In addition to the Polyakov loops, which signal screening by a medium, the observable of central interest
	 is the baryon density 
	\begin{align}
		n_B &= \left.\frac{\partial}{\partial \mu_B} \frac{T}{V}\ln\left(Z\right)\right\vert_{T,V},
	\end{align}
	where $V$ is the (physical) volume.
	
In the regime of very low temperatures we need large values of $N_\tau =100- 500$, which cause the integrands in the partition function defined by the effective action \eqref{eq:Eff_quark_action}
to be very strongly peaked, see also section \ref{sec:CA}. As a result, the amf approach, which is the simplest to evaluate, becomes quantitatively reliable in  this parameter range,
and all our following results are based on it. We have explicitly checked for individual parameter sets, that the numerical results agrees with those of the much 
more expensive rmf method. 
	
	\subsubsection{Baryon onset for heavy quarks\label{sec:Baryon_onset}}
	
	\begin{figure}[!t]\centering
			\includegraphics[width=0.48\textwidth,page=1]{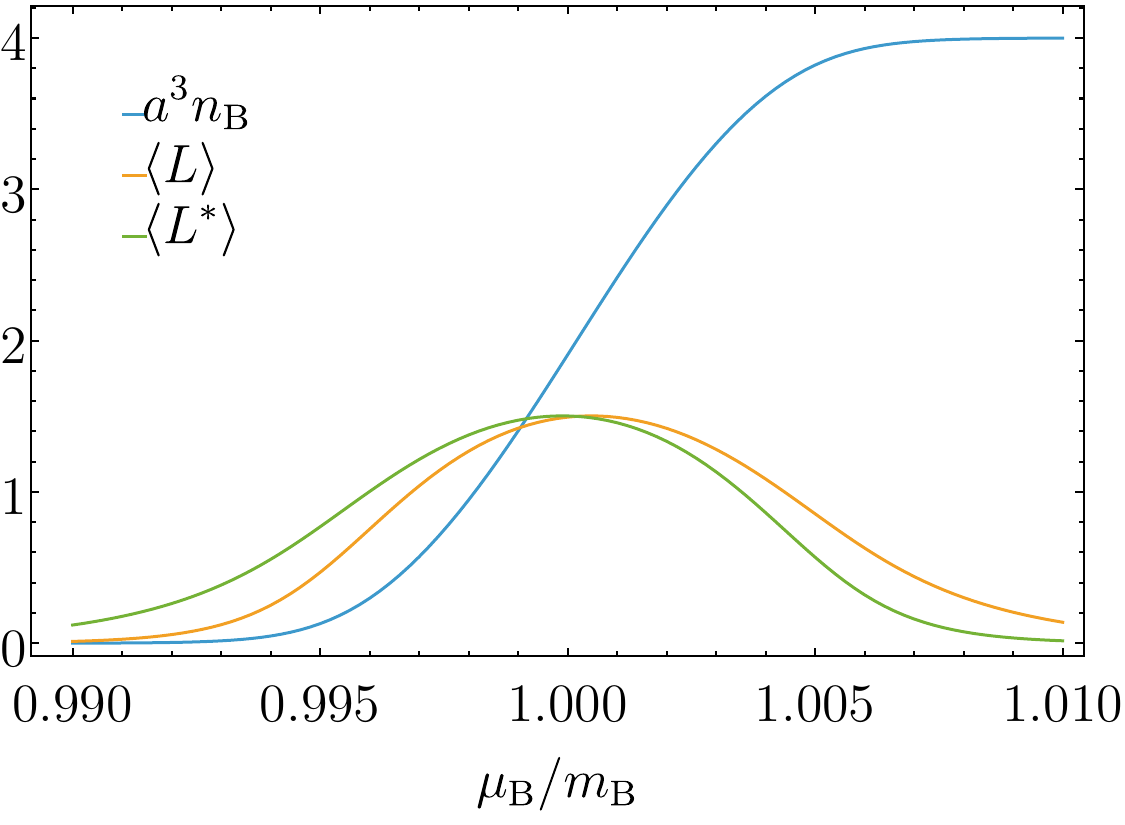}
			\caption{Baryon density $a^3n_B$ as well as expectation values of the Polyakov loop $\langle L \rangle$ and its conjugate $\langle L^* \rangle$. Results are obtained using the amf approach at bare lattice parameters $\kappa = 0.01$, $\beta = 5.7$ and $N_\tau = 100$ as a function of the reduced baryon chemical potential $\mu_B/m_B$. These bare parameters correspond to $m_B \approx 13.59\text{ GeV}$ and $T/m_B \approx 8.52 \times 10^{-4}$.}
			\label{fig:Finite_MuB_Crossover1}
	\end{figure}
	As a first case of interest we consider the regime of the baryon onset within the effective theory using the amf approach at bare parameters $\kappa = 0.01, \beta = 5.7$ and $N_\tau = 100$, corresponding to  $m_B \approx 13.59\text{ GeV}$ and $T/m_B \approx 8.52 \times 10^{-4}$.

	Results for the baryon density as well as the expectation values of the Polyakov loop $\langle L \rangle$ and its conjugate $\langle L^*\rangle$ 	
	are shown in figure \ref{fig:Finite_MuB_Crossover1}.  As expected, all observables start near zero at small $\mu_B$ and increase upon approaching $\mu_B \approx m_B$. Additionally, the Polyakov loops expectation values satisfy $\langle L^* \rangle > \langle L \rangle$ -- a consequence of the conjugated Polyakov loop, representing a static anti-quark, being energetically less costly to excite than the (non-conjugated) Polyakov loop, representing a static quark, due to the abundance of quarks over anti-quarks. Beyond the baryon onset, at even larger $\mu_B$  the baryon density $a^3 n_B$ quickly approaches its saturation level of $2N_f$ per lattice site, due to the Pauli principle. Similarly, since the Polyakov loop represents the colour orientation of a static quark, it is forced to populate all center sectors as the lattice fills up, leading to the vanishing of $\langle L \rangle$ and $\langle L^*\rangle$ in the saturated regime. These results are fully consistent with previous perturbative and numerical approaches to the effective theories \cite{Langelage:2014vpa}. As also observed there, the transition to  a regime with net baryon matter proceeds by a smooth crossover, instead of a first-order liquid-gas transition
	as in physical QCD. This is due to the fact that here we 
	are considering extremely heavy pions, so that the interaction range of the interbaryon Yukawa potential is very short and the binding 
	energy per baryon, which sets the scale for the critical endpoint of the first-order transition, is exponentially small.     
		
	\subsubsection{A first-order nuclear liquid-gas transition \label{sec:eval_liquid_gas}}
	
	\begin{figure}[!t]\centering
		\begin{subfigure}[t]{0.48\textwidth}
			\centering
			\includegraphics[width=\linewidth,page=1]{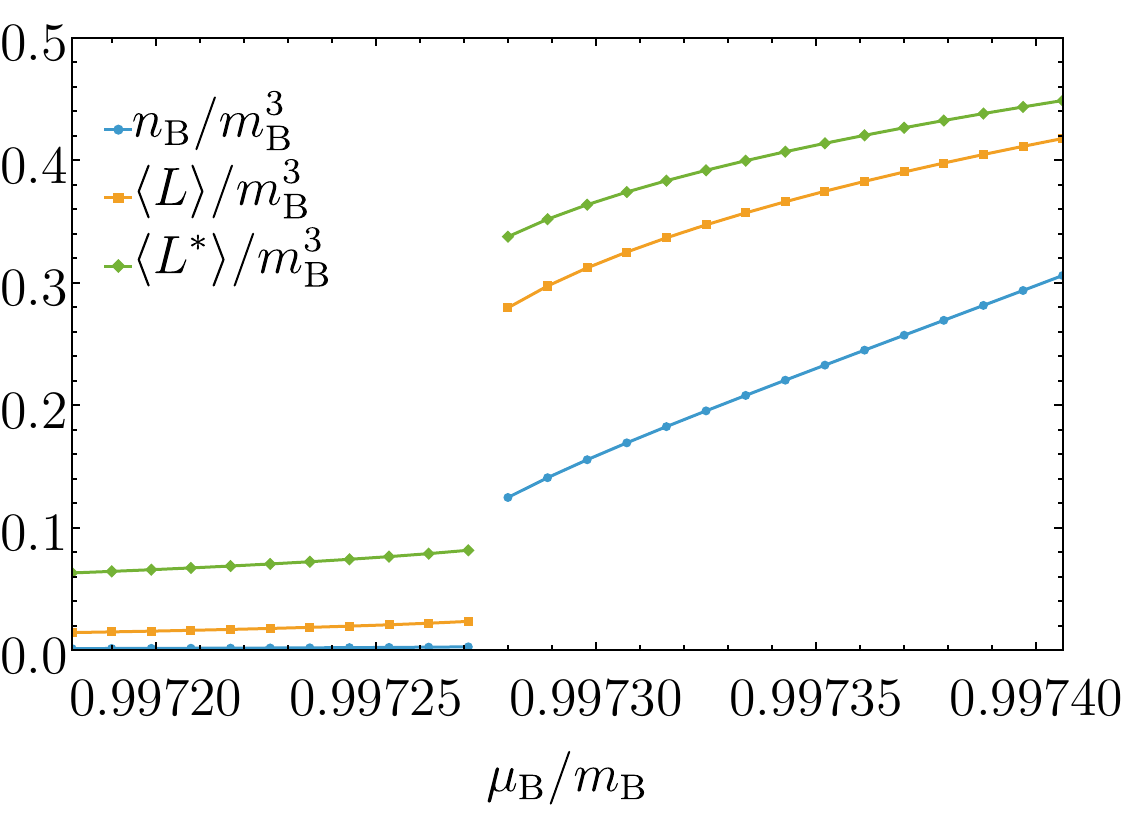}
			\caption{Baryon density $a^3n_B$ as well as expectation values of the Polyakov loops $\langle L \rangle$, $\langle L^* \rangle$. The temperature is here set to $T/m_B \approx 4.88 \times 10^{-4}$, corresponding to bare parameters $\kappa = 0.12$ and $\beta = 5.7$.}
			\label{fig:Finite_MuB_jump}
		\end{subfigure}
		\quad
		\begin{subfigure}[t]{0.48\textwidth}
			\centering
					\includegraphics[width=\linewidth,page=1]{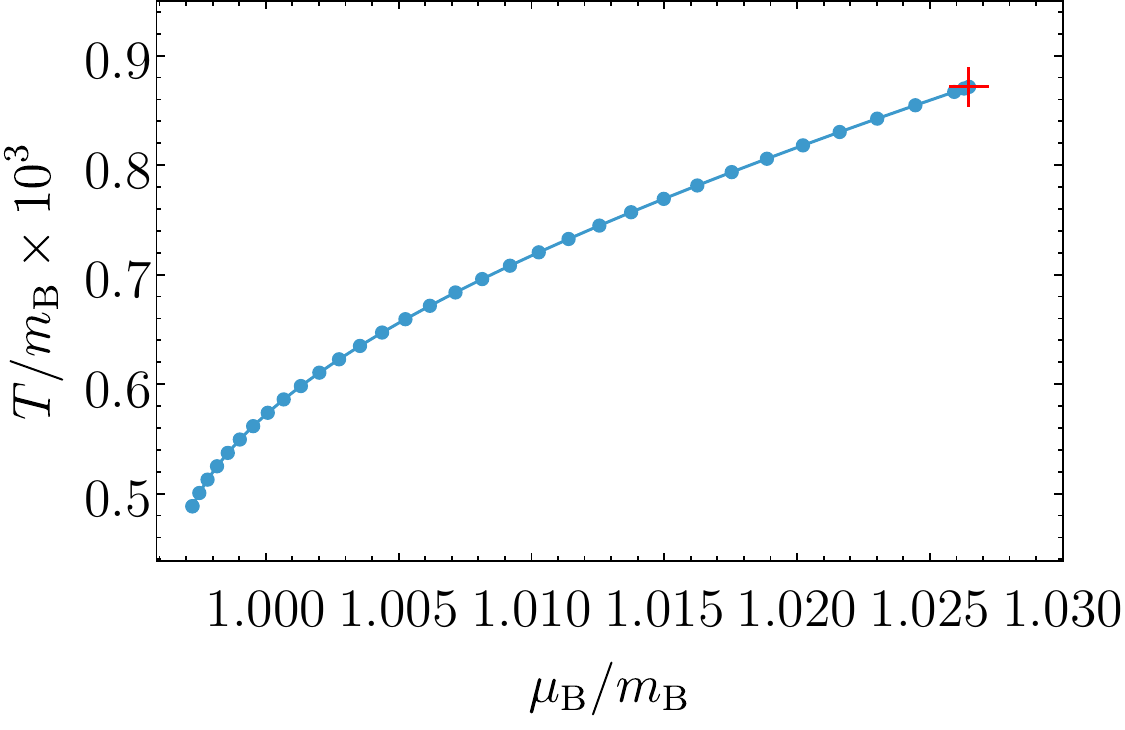}
			\caption{Critical reduced temperature $T_c/m_B$ of the nuclear liquid-gas transition versus $\mu_B/m_B$. The lowest critical temperature is here $T_c/m_B \approx 4.88\times 10^{-4}$, i.e., the same as in figure~\ref{fig:Finite_MuB_jump}. }
			\label{fig:Finite_MuB_crit_line}
		\end{subfigure}
		\caption{Observables and first-order line of the nuclear liquid-gas transition obtained using the amf  approach. Parameters are $N_\tau = 500$, $m_B \approx 4.74\text{ GeV}$ and $N_f = 2$. }
	\end{figure}
	
	To verify this proposition we next consider more moderate baryon masses $m_B\approx 4.74$ GeV, combined with even lower temperatures $T/m_B \approx 4.88\times 10^{-4}$. The corresponding bare lattice parameters are $\kappa = 0.12, \beta = 5.7, N_\tau = 500$. For those bare parameter values a first-order liquid-gas transition was observed numerically in an earlier work~\cite{Langelage:2014vpa}, but with an effective action containing fewer gauge corrections than the present one. Note that a perturbative analysis is hopeless
	in this parameter range because of the size of the coupling $h_2=\kappa N_\tau/N_c$.

A first-order behaviour is also exhibited in our mean-field analysis of the Polyakov loops, $\langle L \rangle_{\text{amf}}$, $\langle L^* \rangle_{\text{amf}}$ and the baryon density, as shown in figure \ref{fig:Finite_MuB_jump}. The expectation values of the Polyakov loops in the gaseous phase are small but non-zero and satisfy $\langle L^* \rangle > \langle L \rangle$, in agreement with section \ref{sec:Baryon_onset}. At a critical $\mu_{B,c}/m_B \approx 0.99726$ both expectation values discontinuously jump towards a larger value, as expected for a first-order nuclear liquid-gas transition. The baryon density in the gaseous phase is nearly zero and jumps at the transition to a positive value, indicating the formation of a medium.

	Using our mean-field approach we can now investigate the behavior of the critical temperature as a function of baryon chemical potential, i.e., map out the first-order line.
	The result is shown in figure \ref{fig:Finite_MuB_crit_line}. One finds the critical temperature $T_c$ to increase with $\mu_B$ until it terminates in a critical end point. While the latter observation is in accord with expectations for a liquid-gas transition, the direction of the line's bending is at odds with physical expectations.
	For physical QCD
	 the Clausius-Clapeyron relation,
	\begin{align}
		\frac{\dif{}}{\dif{\mu_B}}T_c(\mu_B) = -\frac{\Delta n_B}{\Delta s}\label{eq:Clausius_Clapeyron},
	\end{align}	 
	  and the third law of thermodynamics dictate the slope of the transition line to be negative, and it should hit the $T=0$-line at a right angle.
	Here  $\Delta n_B =  n_{B,\text{gas}}- n_{B,\text{liquid}}$ and $\Delta s =  s_{\text{gas}}- s_{\text{liquid}}$ are the differences of the baryon and entropy densities evaluated in the two phases. For physical QCD one expects $\Delta n_B<0$ and 
	$\Delta s <0$\footnote{The entropy per particle is larger in the gaseous phase, but the entropy per volume is lower due to the much higher particle density in the liquid.} across
	the transition, and $\Delta s \rightarrow 0$ for $T\rightarrow 0$. By contrast, our evaluation of the effective theory shows a ``backbending'' of the transition line,
	similar to what is also observed in functional analyses of the quark meson \cite{Tripolt:2017zgc} or baryon meson \cite{Weyrich:2015hha} models. In order
	to understand the origin of this effect, we now proceed to evaluate the entropy density in our effective theories.

	\subsection{The entropy and a systematic error of the effective theories \label{sec:sys}}
	
The entropy density per volume is defined as	
\begin{align}
s &= \left.\frac{\partial}{\partial T} \frac{T}{V}\ln\left(Z\right)\right\vert_{V,\mu_B}.
\end{align}
It has not been calculated explicitly in previous works on the effective theories, but contains valuable information on their validity, as we shall see shortly.
Note that we work in the thermodynamic limit, where $\ln(Z)/V$ and all intensive quantities have no $V$-dependence and result in finite numbers.
	Taking the derivative with respect to the temperature requires some care. This is because for fixed $N_\tau$ the temperature is varied via the lattice spacing, $T=(aN_\tau)^{-1}$, and hence via a change of the bare parameters $\beta$ and $\kappa$ while keeping the baryon mass constant,	
	\begin{align}
	\frac{\partial}{\partial T} =-a^2 N_\tau \Big(\frac{\partial \beta}{\partial a} \frac{\partial}{\partial \beta} + \frac{\partial \kappa}{\partial a} \frac{\partial}{\partial \kappa}\Big).
	\label{eq:chain}
	\end{align}
%	This is achieved by using the implicit function theorem which determines the derivatives of $\beta(a,m_B,N_\tau)$ and $\kappa(a,m_B,N_\tau)$ at fixed $m_B$ and $N_\tau$.
	Since the dynamical change across the nuclear liquid-gas transition is driven by baryons, we simplify our analysis by considering the strong couling limit, $\beta=0$,
	which switches off the gluon dynamics. In this special case we can drop the first term in equation (\ref{eq:chain}).

\begin{figure}[!t]\centering
			\centering
				\begin{subfigure}[t]{0.48\textwidth}
					\centering
			\includegraphics[width=\textwidth,page=1]{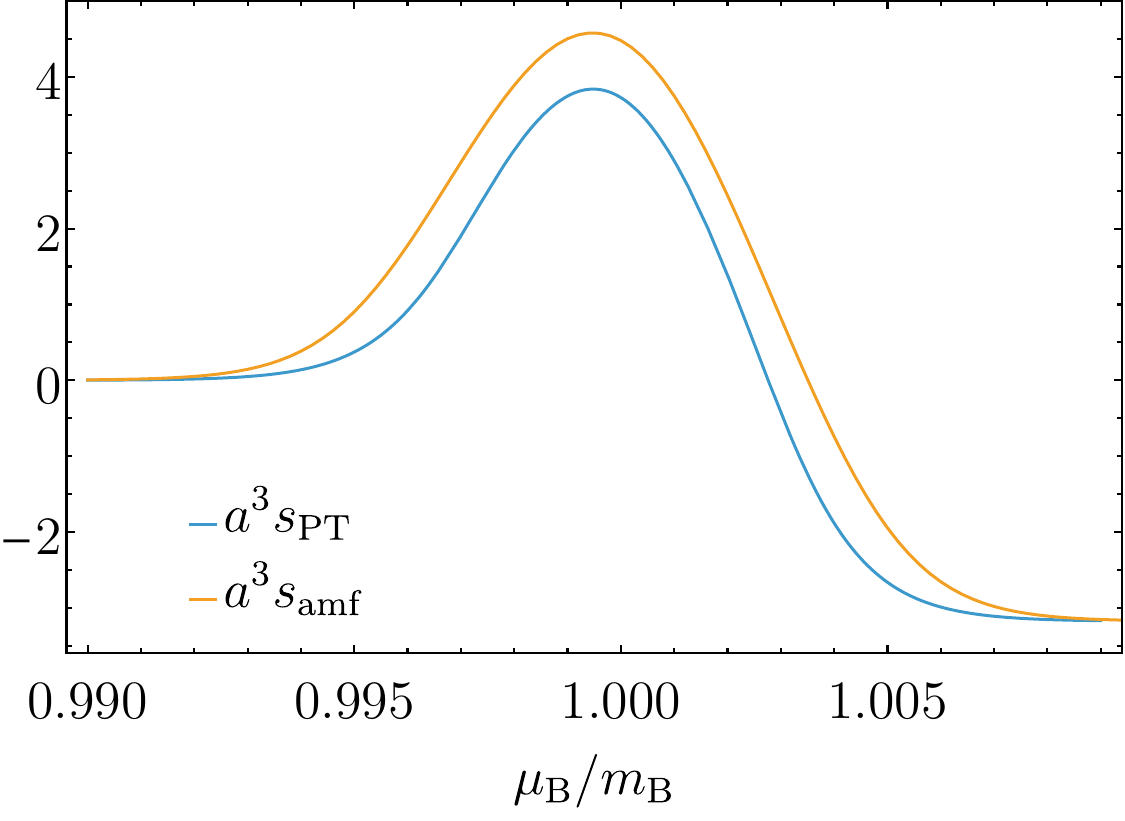}
			\caption{$N_\tau = 100$}
			\end{subfigure}
			\quad
			\begin{subfigure}[t]{0.48\textwidth}
				\centering
				\includegraphics[width=\textwidth,page=1]{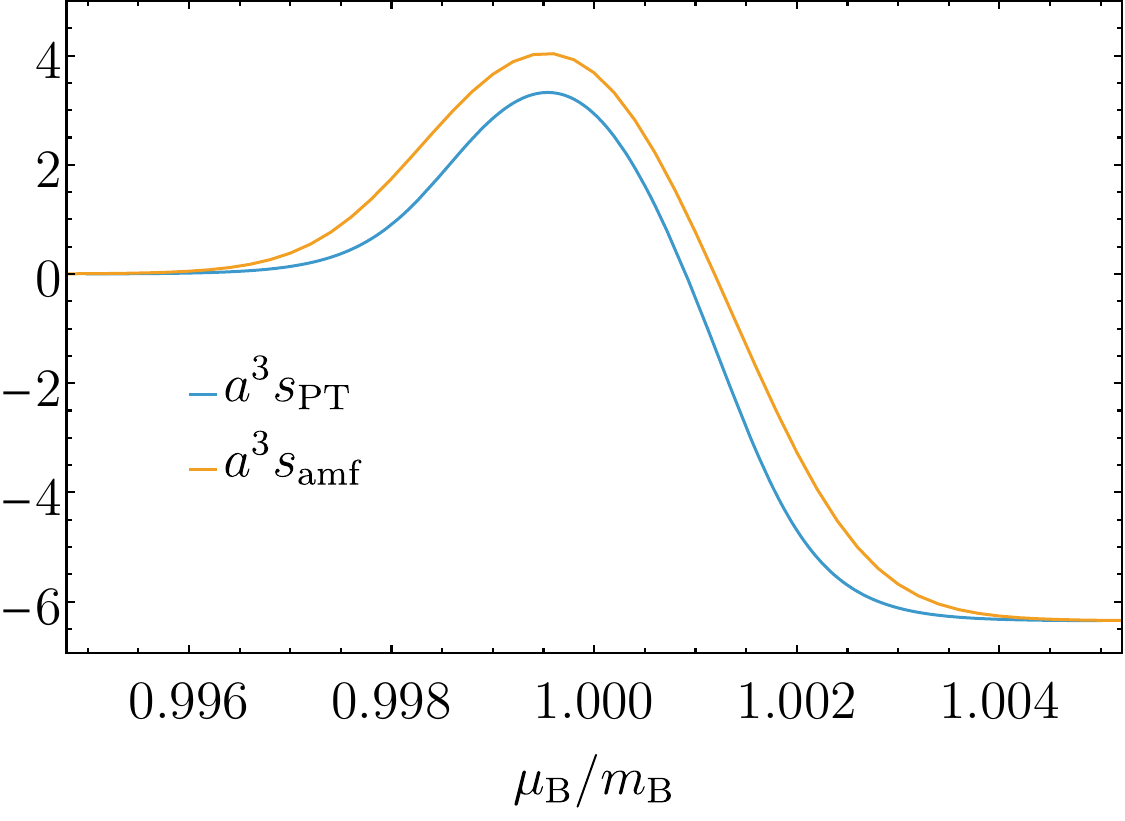}
				\caption{$N_\tau = 200$}
			\end{subfigure}
		\caption{ Entropy densities determined pertubatively (blue line) and via the amf approach (orange line) as a function of the reduced baryon chemical potential $\mu_B/m_B$. Bare parameters are $\kappa = 0.01$ and $\beta = 0$.}
		\label{fig:Finite_MuB_entropy_density-perturbative}
	\end{figure}
Since we are interested in low temperatures, i.e., large $N_\tau$, we evaluate the entropy density within the amf mean-field approach applied to the effective action up to and including 
 $\mathcal{O}(\kappa^2)$ hopping corrections
\eqref{eq:LOKineticAction}\footnote{
Within this effective action, the only contribution to the free energy density in the vacuum comes from the volume of the Haar-measure, which is normalized to $1$. The amf approach, however, violates this normalization and instead gives a group volume of $27/6 = 9/2$, see section \ref{sec:CA}. This constant has to be subtracted to get a vacuum entropy of zero also in the amf mean-field approach.}.
 In order to test for consistency, we 
compare with an earlier perturbative analysis \cite{Langelage:2014vpa} of pressure $p$, number density $n_B$ and energy density $e$ to $\mathcal{O}(\kappa^2)$ at $\beta = 0$, using the fundamental relation $-p = e - Ts -\mu_B n_B$, which leads to
\begin{align}
a^3s= N_\tau(a^4e+a^4p-a\mu_B\;a^3n_B).
\label{eq:pert_s}
\end{align}
We also  employ the leading hopping approximation for the baryon mass in lattice units, which is accurate up to and including $O(\kappa^2)$ in the strong-coupling limit,
\begin{align}
	am_B(\kappa) = -3\log(2\kappa) + \mathcal{O}(\kappa^3),
\end{align}
giving a simple relation between $\kappa$ and the lattice spacing $a$ at fixed baryon mass $m_B$.
 The resulting perturbative entropy density is  explicitly given by
	\begin{align}
	a^3s=&	h_2 \left(-\frac{24 h_1 z_1^2 z_2 \log (h_1)}{z_0^3}+\frac{24 h_1 z_1 z_3 \log (h_1)}{z_0^2}+\frac{24 z_1^2 \log (2 k)}{z_0^2}-\frac{12 z_1^2}{z_0^2}\right)\notag\\
	&-\frac{h_1 z_2 \log
		(h_1)}{z_0}+\log (z_0) + \mathcal{O}(\kappa^4) \label{eq:Entropy_pert_expression}\\
	z_0 =&  1 + 20 h_1^3 + 50 h_1^6 + 20 h_1^9 + h_1^{12} \\
	z_1 =& 15 h_1^3 + 75 h_1^6 + 45 h_1^9 + 3 h_1^{12} \\
	z_2 =& 60 h_1^2 + 300 h_1^5 + 180 h_1^8 + 12 h_1^{11} \\
	z_3 =& 45 h_1^2 + 450 h_1^5 + 405 h_1^8 + 36 h_1^{11},
	\end{align}
	with the leading expression for the effective coupling $h_2=\kappa^2N_\tau/N_c$.  In order to stay in the perturbative regime, we choose a sufficiently small $\kappa N_\tau$-value,
	where the transition proceeds by crossover.  
	
	Figure~\ref{fig:Finite_MuB_entropy_density-perturbative} compares this perturbative result (blue lines) with the entropy density obtained via the amf mean-field approach (orange lines). Note that the temperature is not exactly zero as long as $N_\tau$ is finite. In the baryon onset region, $\mu_B \approx m_B$, the entropy densities increase as the lattice gets populated, in accord with a growing number of baryon states becoming available for the system compared to the vacuum. When the lattice approaches saturation restrictions on the available states from the Pauli principle become increasingly relevant, resulting in a reduction of the entropy densities. 
At saturation there is only a single fermionic state available for the system (the one where every lattice site is fully occupied with quarks), while gluons are switched off at $\beta=0$. This must result in zero entropy at lattice saturation. By contrast, we observe an unphysical negative entropy, both in mean field and perturbation theory. Moreover, the saturation level gets more negative with increasing $N_\tau$, see figure \ref{fig:Finite_MuB_entropy_density-perturbative}.
Note that this problem is not visible on the vacuum side, which
correctly reproduces the silver blaze property, due to the exponential suppression caused by $h_1\sim \exp(-\mathrm{const} \cdot N_\tau)$.

We then have  two systematic errors working in the same direction: the fact that the lattice saturates as well as the associated negative entropy saturation level both cause a decrease of
the entropy compared to its true level on the liquid side of the phase transition, thus effectively reversing the sign of $\Delta s$ and causing the incorrect slope and curvature
of the first-order transition line.

The saturation limit of the perturbative entropy density is determined by the large density limit $\mu_B\to\infty$ (or equivalently $h_1 \to \infty$) of  \eqref{eq:Entropy_pert_expression}, 
\begin{align}
	a^3 s_{\text{PT}}(\mu_B \to \infty) =& -108 h_2 (1 - 2 \ln(2 \kappa)) + \mathcal{O}(\kappa^4).
\end{align}
Within mean field it is also possible to determine the saturation value of the entropy density analytically, as the field dependence of the effective action  \eqref{eq:LOKineticAction} cancels in the $\mu_B\to\infty$ limit, resulting in the same expression,
\begin{align}
	a^3 s_{\text{CA}}(\mu_B \to \infty) =& -108 h_2 (1 - 2 \ln(2 \kappa)) + \mathcal{O}(\kappa^4).
\end{align}
The reason for the negative saturation level is now quickly identified to be the term proportional to the nearest-neighbour coupling, $h_2=\kappa^2N_\tau/N_c$,
which diverges at fixed lattice spacing with $N_\tau\rightarrow \infty$, leading to $s(\mu_B > m_B)\rightarrow -\infty$ as $N_\tau \rightarrow \infty$. 
%Besides prohibiting a perturbative evaluation of the effective theory for large values of $N_\tau$, 
Such a functional
behaviour is clearly unphysical and represents an artefact of the truncated hopping series, affecting both perturbative and non-perturbative studies of the effective theories at large $N_\tau$. To understand this point, it is instructive to compare with the resummed high-order expression 
for the effective gauge coupling, (\ref{eq:Lambda1Expression}). Expanded and truncated at next-to-leading order, this reads
\begin{align}
\lambda_1(u)=u^{N_\tau}\Big( 1 + 4 N_\tau\;u^4+\ldots \Big),
\end{align}
apparently containing a factor linear in $N_\tau$. Once properly combined with higher orders, however, the functional behaviour is different. We similarly expect
the $N_\tau$ dependence of $h_2$ as well as all $h_{3_i}$ (see appendix \ref{sec:qeff}) to be fundamentally modified
once sufficiently high orders in $\kappa$ are
available to recognise appropriate all-order resummations. We suspect this truncated $N_\tau$-dependence of $h_2$ to also be at the heart of the
deviation of the deconfinement critical endpoint from the simulations of full QCD with increasing $N_\tau$, as discussed in section \ref{subsec:eval_decon}.

	\section{Conclusions}
	
	In this work mean-field techniques were applied to dimensionally reduced effective Polyakov loop theories of LQCD that have been derived in earlier works via combined strong-coupling and hopping parameter expansions. These effective theories are known to have a significantly reduced sign problem in numerical simulations 
	%even at $\mu_B\not = 0$ than full LQCD. Thus, in previous works these effective theories 
	and were used to investigate the deconfinement transition at $\mu_B \ge 0$ \cite{Fromm2013} as well as the low-temperature finite-density regime of the QCD phase diagram. In the latter evidence for a first-order nuclear liquid-gas transition was found via simulations in small volumes \cite{Langelage:2014vpa}. 
	
	Here, we introduced three variations of mean-field approximations including fluctuations to a different degree, and discussed their application at $\mu_B \not = 0$. As a benchmark we first determined the critical couplings for the first-order deconfinement transition in the pure gauge limit. While the least accurate mean-field approach yields only qualitative results (about 50\% relative error) the most accurate approach reproduces Monte Carlo results at percent level. For QCD with heavy quarks we followed the deconfinement line up to its critical end point 
	at some critical hopping parameter, viz.~quark mass, whose location is however not accurate, as expected for a mean-field approach. On the other hand, the deconfinement
	transition and its critical endpoint can readily be determined also at fixed quark mass and
	 $\mu_B \ge 0$, for which we point out the difference between reweighted phase-quenched and full mean-field treatments. 
	
	In the cold and dense regime our mean-field approach permits also analytical studies of the first-order nuclear liquid-gas transition for sufficiently light quarks, in qualitative agreement with previous numerical \cite{Fromm2013} results. On the liquid side of this transition we found unphysical behaviour for the entropy density in the zero
	temperature and lattice saturation limits, whose cause we identified to be the truncation of the hopping expansion in the nearest-neighbour Polyakov loop interaction.
	We expect the same issue to be at the root of the unexpected continuum approach of the deconfinement critical end point, and hope to cure it	
 by devising an appropriate hopping resummation in future work.
 	
	\section*{Acknowledgments}
	We thank Jonas Scheunert for collaboration in the early stages of this project. The authors acknowledge support by the Deutsche Forschungsgemeinschaft (DFG, German Research Foundation) through the CRC-TR 211 'Strong-interaction matter under extreme conditions'- project number 315477589 - TRR 211 and by the State of Hesse within the Research Cluster ELEMENTS (Project ID 500/10.006).
	
	\appendix
	\section{The effective action used within this work \label{sec:DetailsEffectiveActionUsed}}
	\subsection{pure gauge effective coupling \label{sec:DiscussionLambda1Expression}}
	The expression of the pure gauge effective coupling $\lambda_1(\beta, N_\tau)$ used within this work has been derived for $N_c = 3$ and $N_\tau \ge 4$ in \cite{Langelage2011}. It is expressed in terms of a resummed series in the character expansion coefficient of the fundamental representation $u$,
	\begin{align}
		 u(\beta) :=& \frac{c_f(\beta)}{3 c_0(\beta)},\quad c_r := \int\dif{U}\chi_r(U^\dagger)\exp\left(\frac{\beta}{3}\ReN\tr\left[1-U\right]\right),
	\end{align}
	where $\chi_r(\cdot)$ is the character of the irreducible representation $r$ of the gauge group. For the coefficients $c_f$ and $c_0$ we employed the series expansions
	\begin{align}
		 c_f(\beta) :=& \frac{\beta}{6}+\frac{1}{2}\left(\frac{\beta}{6}\right)^2+\left(\frac{\beta}{6}\right)^3+\frac{5}{8} \left(\frac{\beta}{6}\right)^4+\frac{13 }{24}\left(\frac{\beta}{6}\right)^5+\frac{77}{240} \left(\frac{\beta}{6}\right)^6+\frac{139 }{720}\left(\frac{\beta}{6}\right)^7\notag\\
		 &+\frac{19 }{192}\left(\frac{\beta}{6}\right)^8+\frac{23 }{480}\left(\frac{\beta}{6}\right)^9+\frac{319 }{15120}\left(\frac{\beta}{6}\right)^{10}+\frac{2629
		 	}{302400}\left(\frac{\beta}{6}\right)^{11}+\frac{16133 }{4838400}\left(\frac{\beta}{6}\right)^{12}\notag\\
		 	&+\frac{17449 }{14515200}\left(\frac{\beta}{6}\right)^{13}+\frac{35531 }{87091200}\left(\frac{\beta}{6}\right)^{14}+\frac{79837 }{609638400}\left(\frac{\beta}{6}\right)^{15}+\frac{1457971
		 }{36578304000}	\left(\frac{\beta}{6}\right)^{16}\notag\\
		 &+\frac{10421 }{903168000}\left(\frac{\beta}{6}\right)^{17}+\frac{569411 }{178827264000}\left(\frac{\beta}{6}\right)^{18}+\frac{11261 }{13412044800}\left(\frac{\beta}{6}\right)^{19}+\mathcal{O}\left(\beta^{20}\right)
		\end{align}
		\begin{align}
		 c_0(\beta) :=& 1+\left(\frac{\beta}{6}\right)^2+\frac{1}{3}\left(\frac{\beta}{6}\right)^3+\frac{1}{2}\left(\frac{\beta}{6}\right)^4+\frac{1}{4}\left(\frac{\beta}{6}\right)^5+\frac{13 }{72}\left(\frac{\beta}{6}\right)^6+\frac{11 }{120}\left(\frac{\beta}{6}\right)^7\notag\\
		 &+\frac{139 }{2880}\left(\frac{\beta}{6}\right)^8+\frac{19}{864} \left(\frac{\beta}{6}\right)^9+\frac{23 }{2400}\left(\frac{\beta}{6}\right)^{10}+\frac{29 }{7560}\left(\frac{\beta}{6}\right)^{11}+\frac{2629
		 	}{1814400}\left(\frac{\beta}{6}\right)^{12}\notag\\
		 	&+\frac{1241 }{2419200}\left(\frac{\beta}{6}\right)^{13}+\frac{17449 }{101606400}\left(\frac{\beta}{6}\right)^{14}+\frac{35531 }{653184000}\left(\frac{\beta}{6}\right)^{15}+\frac{79837 }{4877107200}\left(\frac{\beta}{6}\right)^{16}\notag\\
		 	&+\frac{85763
		 	}{18289152000}\left(\frac{\beta}{6}\right)^{17}+\frac{10421 }{8128512000}\left(\frac{\beta}{6}\right)^{18}+\frac{29969 }{89413632000}\left(\frac{\beta}{6}\right)^{19}+\mathcal{O}\left(\beta^{20}\right),
	\end{align}
	which extend the known expressions from $\mathcal{O}(\beta^{14})$ \cite{Montvay1994} to $\mathcal{O}(\beta^{19})$. The resummed expression for the effective coupling used within this work is\cite{Langelage2011}
	\begin{align}
		\lambda_1(u(\beta), N_\tau = 4) =& u^{N_\tau} \exp\left[N_\tau\left(4 u^4 + 12 u^5 - 14 u^6 - 36 u^7 + \frac{295}{2} u^8 + \frac{1851}{10} u^9 \right.\right.\notag\\
		&\left.\left.+ 
		\frac{1035317}{5120} u^{10} + \mathcal{O}(u^{11})\right)\right]\notag\\
		\lambda_1(u(\beta), N_\tau > 4) =& u^{N_\tau} \exp\left[N_\tau\left(4 u^4 + 12 u^5 - 14 u^6 - 36 u^7 + \frac{295}{2} u^8 + \frac{1851}{10} u^9 \right.\right.\notag\\
		&\left.\left.+ 
		\frac{1055797}{5120 }u^{10} + \mathcal{O}(u^{11})\right)\right] \label{eq:Lambda1Expression}.
	\end{align}
	
	\subsection{The quark effective action \label{sec:qeff}}
	
	We split the quark effective action $\quarkEffAction$ used within this work into interactions,
	\begin{align}
		-\quarkEffAction =:& N_f\ln \left(\det Q_\text{S}\right) + \sum_{\langle \mathbf{x},\mathbf{y}\rangle} I_{\text{K,nn}}\left(W_{\mathbf{x}}, W_{\mathbf{y}}\right) + \sum_{(\mathbf{x}, \mathbf{y}, \mathbf{z}) = \lrcorner}I_{\text{K,c}}\left(W_{\mathbf{x}}, W_{\mathbf{y}}, W_{\mathbf{z}}\right) \notag \\ 
		&+ \sum_{(\mathbf{x}, \mathbf{y}, \mathbf{z})=\text{---}}I_{\text{K,s}}\left(W_{\mathbf{x}}, W_{\mathbf{y}}, W_{\mathbf{z}}\right), \label{eq:Eff_quark_action}
	\end{align}
	where $I_{\text{K},\text{nn}}$ contains the nearest-neighbor interaction terms, while $I_{\text{K,c}}$ describes the three point interactions forming a corner, and $I_{\text{K,s}}$ describes the three point interaction along a straight line. The sum over $(\mathbf{x}, \mathbf{y}, \mathbf{z}) = \lrcorner$ ($(\mathbf{x}, \mathbf{y}, \mathbf{z})=\text{---}$) corresponds to a summation over all corners (straight lines) on the lattice, with midpoint $\mathbf{y}$ and end points $\mathbf{x}$ and  $\mathbf{z}$.  In this work we used expressions for these three interactions correct to $\mathcal{O}(\kappa^4)$, as derived in \cite{Langelage:2014vpa}. The nearest-neighbour interaction is
	\begin{align}
		I_{\text{K},\text{nn}}\left(W_{\mathbf{x}}, W_{\mathbf{y}}\right) =& -2N_f h_2 W_{1111}^{-}(W_{\mathbf{x}})W_{1111}^{-}(W_{\mathbf{y}}) - \frac{1}{2}\left(2N_f h_2 W_{1111}^{-}(W_{\mathbf{x}})W_{1111}^{-}(W_{\mathbf{y}})\right)^2 \notag\\
		&+N_f h_{3_1}\left[
		\left(W^+_{2121}(W_{\mathbf{x}}) + \frac{2}{N_\tau - 1}\frac{(2\kappa)^2 - (2\kappa)^{2N_\tau}}{1 - (2\kappa)^2}W_{1010}(W_{\mathbf{x}})\right) W_{1111}^{-}(W_{\mathbf{y}})^2 \right.\notag\\
		&+\left. W_{1111}^{-}(W_{\mathbf{x}})^2 \left(W^+_{2121}(W_{\mathbf{y}}) + \frac{2}{N_\tau - 1}\frac{(2\kappa)^2 - (2\kappa)^{2N_\tau}}{1 - (2\kappa)^2}W_{1010}(W_{\mathbf{y}})\right)
		\right] \notag\\
		&-N_f h_{3_2}\frac{N_c^2}{N_c^2 - 1} \left[\vphantom{\frac{1}{1}}\left(W^+_{2222}(W_{\mathbf{x}}) - 2W_{1111}(W_{\mathbf{x}})\right)W_{1111}^{-}(W_{\mathbf{y}})^2 \right. \notag\\
		&\left.+W_{1111}^{-}(W_{\mathbf{x}})^2 \left(W^+_{2222}(W_{\mathbf{y}}) - 2W_{1111}(W_{\mathbf{y}})\right)
		\vphantom{\frac{1}{1}}\right] \notag\\
		&+N_f h_{3_2}\frac{N_c}{N_c^2 - 1}\left[\vphantom{\frac{1}{1}}
		\left(W^+_{2222}(W_{\mathbf{x}}) - 2W_{1111}(W_{\mathbf{x}})\right)\left(W^+_{2222}(W_{\mathbf{y}}) - 2W_{1111}(W_{\mathbf{y}})\right) \right. \notag\\
		&\left.+ W_{1111}^{-}(W_{\mathbf{x}})^2 W_{1111}^{-}(W_{\mathbf{y}})^2
		\vphantom{\frac{1}{1}}\right]\notag\\
		&+2N_f^2 h_{3_1}\left[\vphantom{\frac{1}{1}}
		W_{1111}^{-}(W_{\mathbf{x}})^2 W_{1111}^{-}(W_{\mathbf{y}})^2 + W_{2121}^{+}(W_{\mathbf{x}}) W_{2121}^{+}(W_{\mathbf{y}}) \right.\notag\\
		&\left.+ \frac{2}{N_\tau -1}\frac{(2\kappa)^2 - (2\kappa)^{2N_\tau}}{1-(2\kappa)^2} \left(W_{2121}^{+}(W_{\mathbf{x}}) W_{1010}(W_{\mathbf{y}}) + W_{1010}(W_{\mathbf{x}})W_{2121}^{+}(W_{\mathbf{y}}) \right) \right.\notag\\
		&\left.+ \left(\frac{2}{N_\tau-1}\frac{(2\kappa)^4 - (2\kappa)^{4N_\tau}}{1-(2\kappa)^4}+ 2(2\kappa)^{2N_\tau}\right)W_{1010}(W_{\mathbf{x}})W_{1010}(W_{\mathbf{y}}) 
		\vphantom{\frac{1}{1}}\right] \notag\\
		&+2N_f^2h_{3_2}\frac{N_c^2}{N_c^2 - 1}\left[\vphantom{\frac{1}{1}}
		W_{1111}^{-}(W_{\mathbf{x}})^2 W_{1111}^{-}(W_{\mathbf{y}})^2 \right.\notag\\
		&\left.+ \left(W^+_{2222}(W_{\mathbf{x}}) - 2W_{1111}(W_{\mathbf{x}})\right)\left(W^+_{2222}(W_{\mathbf{y}}) - 2W_{1111}(W_{\mathbf{y}})\right)
		\vphantom{\frac{1}{1}}\right] \notag\\
		&-2N_f^2h_{3_2}\frac{N_c}{N_c^2-1}\left[\vphantom{\frac{1}{1}}\left(W^+_{2222}(W_{\mathbf{x}}) - 2W_{1111}(W_{\mathbf{x}})\right) W_{1111}^{-}(W_{\mathbf{y}})^2 \right.\notag\\
		&\left.+ W_{1111}^{-}(W_{\mathbf{x}})^2\left(W^+_{2222}(W_{\mathbf{y}}) - 2W_{1111}(W_{\mathbf{y}})\right)\vphantom{\frac{1}{1}}\right] + \mathcal{O}(\kappa^6).
	\end{align}
	The effective interaction for three points forming a corner is given by
	\begin{align}
		I_{\text{K,c}}\left(W_{\mathbf{x}}, W_{\mathbf{y}}, W_{\mathbf{z}}\right) =& 2N_f h_{3_3} \frac{N_\tau - 1}{N_\tau} W^{-}_{1111}(W_{\mathbf{x}})W^{+}_{2121}(W_{\mathbf{y}})W^{-}_{1111}(W_{\mathbf{z}})\notag\\
		&-N_f h_{3_3}\frac{1}{N_\tau} W^{-}_{1111}(W_{\mathbf{x}}) \left(4N_c  + 2W^+_{2222}(W_{\mathbf{y}}) - 4W^+_{1111}(W_{\mathbf{y}})\right) \notag\\
		&\times  W^{-}_{1111}(W_{\mathbf{z}}) + \mathcal{O}(\kappa^6),
	\end{align}
	whereas the effective interaction for the three point straight line is
	\begin{align}
		I_{\text{K,s}}\left(W_{\mathbf{x}}, W_{\mathbf{y}}, W_{\mathbf{z}}\right) =& 2N_f h_{3_3}\frac{N_\tau - 1}{N_\tau} W^{-}_{1111}(W_{\mathbf{x}}) \left(\vphantom{\frac{1}{1}}W^+_{2121}(W_{\mathbf{y}}) \right. \notag\\
		&\left.- \frac{2}{N_\tau -1}\frac{(2\kappa)^2 - (2\kappa)^{2N_\tau}}{1-(2\kappa)^2}W_{1010}(W_{\mathbf{y}})\right)W^-_{1111}(W_{\mathbf{z}})\notag\\
		&-2N_f h_{3_3}\frac{1}{N_\tau}W^-_{1111}(W_{\mathbf{x}}) \left(4N_c + W^+_{2222}(W_{\mathbf{y}}) - 4W^+_{1111}(W_{\mathbf{y}}) + 2W_{1111}(W_{\mathbf{y}})\right) \notag\\
		&\times W^-_{1111}(W_{\mathbf{z}}) + \mathcal{O}(\kappa^6).
	\end{align}
For the quark effective couplings we employed expressions  that resum gauge corrections as derived in \cite{NeumanPhD}\footnote{Note that the expressions for the effective couplings given here correct typos in \cite{NeumanPhD}.},
	\begin{align}
		h_1 =& \exp\left[N_\tau (a\mu + \ln(2\kappa)) + 6N_\tau \kappa^2 u \left(\frac{1-u^{N_\tau - 1}}{1-u} + 4u^4 - 12\kappa^2 + 9\kappa^2u + 4\kappa^2 u - 4\kappa^4\right)\right]\label{eq:FullExpression:h1}\\
		h_2 =& \frac{N_\tau \kappa^2}{N_c}\left(1 + 2\frac{u-u^{N_\tau}}{1-u} + 8u^5 + 16\kappa^2u^4\right)\label{eq:FullExpression:h2}\\
		h_{3_1} =& \frac{N_\tau(N_\tau - 1)\kappa^4}{N_c^2}\times \begin{cases}
			1 + \frac{8}{3}\left(u + u^2 + 4u^5 + 8\kappa^2 u^4\right),&\text{if } N_\tau = 4\\
			1+\frac{2}{5}\left(8u + 12u^2 + 12u^3 + 8u^4 + 32u^5 + 64\kappa^2u^4\right),&\text{if } N_\tau = 6\\
			1+\frac{8}{7}\left(3u + 5u^2 + 6u^3 + 6u^4 + 17u^5 + 24\kappa^2u^4\right),&\text{if } N_\tau = 8\\
			1+\frac{4u(2-N_\tau + N_\tau u)}{(N_\tau-1)(u-1)^3} + \frac{20(N_\tau - 6)}{N_\tau  - 1}u^5 + \frac{40(N_\tau - 6)}{N_\tau - 1} \kappa^2 u^4,&\text{if } N_\tau > 8\\
		\end{cases}\label{eq:FullExpression:h31}\\
		h_{3_2} =& \frac{N_\tau \kappa^4}{N_c^2}\left(1 + 4\frac{u - u^{N_\tau}}{1-u} + 16u^5 + 32\kappa^2u^4\right)\label{eq:FullExpression:h32}\\
		h_{3_3} =& \frac{N_\tau^2 \kappa^4}{N_c^2}\left(1 + 4\frac{(1-u^{N_\tau})(u-u^{N_\tau})}{(1-u)^2} + 16u^5 + 32\kappa^2 u^4\right)\label{eq:FullExpression:h33}
	\end{align}
	The static anti-quark effective coupling $\bar{h_1}_1$ receives the same corrections as ${h_1}_1$, i.e., we have $\bar{h_1}_1(\mu) = {h_1}_1(-\mu)$. 
	
	Another resummation that was employed in this work absorbs quark propagations along squares on the euclidean lattice into a $\kappa$-dependent shift of $\beta$ for all $u$ that appear in \ref{sec:DetailsEffectiveActionUsed}. This amounts to the replacement \cite{Fromm2012}
	\begin{align}
		u = u(\beta) \rightarrow u(\beta + 48N_f \kappa^4).
	\end{align} 
\section{Eigenvalues of temporal Wilson lines and Polyakov loops \label{sec:Relation_EVs_Polyakov_loops}}
A third option to obtain $W_{nm\bar n \bar m}(L_{\mathbf{x}}, L^*_{\mathbf{x}})$ is to notice that due to the traces in \eqref{fractionalWilsonLoop} the expression only depends on the eigenvalues of $W_{\mathbf{x}}$. By the SU($3$)-constraints $\det(W_{\mathbf{x}}) = 1$ and $W_{\mathbf{x}}W_{\mathbf{x}}^{\dagger} = 1$ there are only two independent eigenvalues, which we now call $z_1$ as well as $z_2$, and both lie on the complex unit circle. The third eigenvalue $z_3$ is simply given by $z_3 = 1/(z_1 z_2)$. The two Polyakov loops $L_{\mathbf{x}}$ and $L^*_{\mathbf{x}}$ are related to these eigenvalues by
\begin{align}
	L_{\mathbf{x}} = z_1 + z_2 + \frac{1}{z_1 z_2}\quad\text{and}\quad L_{\mathbf{x}}^* = \frac{1}{z_1 }+ \frac{1}{z_2} + z_1 z_2.\label{eq:equationSystemEVs}
\end{align}
This system of two equations with the two unknowns $z_1$ and $z_2$ can be solved analytically (e.g. using Mathematica) and determines $z_1(L_{\mathbf{x}},L^*_{\mathbf{x}})$ and $z_2(L_{\mathbf{x}},L^*_{\mathbf{x}})$. The expression for $W_{nm\bar n \bar m}(L_{\mathbf{x}}, L^*_{\mathbf{x}})$ is then simply obtained after performing the substitution
\begin{align}
	W_{\mathbf{x}}\rightarrow \begin{pmatrix}
		z_1(L_{\mathbf{x}},L^*_{\mathbf{x}}) & & \\
		& z_2(L_{\mathbf{x}},L^*_{\mathbf{x}}) &\\
		& & \frac{1}{z_1 z_2(L_{\mathbf{x}},L^*_{\mathbf{x}})}
	\end{pmatrix}
\end{align}
in the left-most equation of \eqref{fractionalWilsonLoop}.

Note that there are six solutions to \eqref{eq:equationSystemEVs} corresponding to the $3!$ ways of permuting the eigenvalues. The effective action only depends on sums over these eigenvalues and is thus independent on the particular choice of solution for $z_1$ and $z_2$. This is consistent with expectations from the Cayley-Hamilton theorem.

For numerical evaluations of the effective theories, where one makes use of the change of variables \eqref{eq:ChangeOfVariables} with \eqref{eq:Change_to_reduced_Haar_measure}, one may simply read off $z_1 = e^{i\phi_1}$ and $z_2 = e^{i\phi_2}$ from \eqref{eq:ChangeOfVariables}.
\bibliographystyle{JHEP}
\bibliography{library}
\end{document}